\documentclass[aps,prb,reprint,groupedaddress]{revtex4-2}

\usepackage{graphicx} 
\usepackage{amsmath} 
\usepackage[colorlinks=true,allcolors=blue]{hyperref} 
\usepackage[capitalise]{cleveref} 

\begin{document}

\title{Symmetric Helmholtz Fermi-surface harmonics for an optimal representation of anisotropic quantities on the Fermi surface: Application to the electron-phonon problem}

\author{Jon Lafuente-Bartolome}
\author{Idoia G. Gurtubay}
\author{Asier Eiguren}
\affiliation{Materia Kondentsatuaren Fisika Saila, University of the Basque Country UPV/EHU, 48080 Bilbao, Basque Country, Spain}
\affiliation{Donostia International Physics Center (DIPC), Paseo Manuel de Lardizabal 4, 20018 Donostia-San Sebasti\'{a}n, Spain}

\date{\today}

\begin{abstract}
We outline a numerical procedure to incorporate the crystal symmetries in the Helmholtz Fermi-surface harmonics basis set,
which are the solutions of the Helmholtz equation defined on the Fermi surface.
This improvement allows for an optimal representation of anisotropic quantities defined on the Fermi surface in terms of few symmetric elements of the set.
We demonstrate the general validity of our approach by identifying the fully symmetric Helmholtz Fermi-surface harmonics subset for several representative systems with different crystal structures, namely, FCC-Cu, HEX-MgB$_2$, and BCC-YH$_6$.
Furthermore, we illustrate the potential of the method applied to the electron-phonon problem, showing that the anisotropic electron-phonon mass-enhancement parameter $\lambda_{\bf k}$ can be represented to high accuracy by a handful of coefficients.
This works as an effective filter, paving the way for a reduction of several orders of magnitude in the computation of superconductivity, impurity problems,  or any other Fermi surface dependent property of metals from first principles.
\end{abstract}


\maketitle


\section{introduction}

Due to the Pauli exclusion principle, the low-energy electronic excitations in a metal are restricted to a very narrow window around its Fermi surface (FS).
Consequently, the transport properties of metals at finite temperatures and/or external electromagnetic fields are governed by the specific shape and topology of its FS, and by the details of the matrix elements defining the scattering processes on it.
One of the most important scattering source for electrons in metals is their interaction with the lattice vibrations or phonons \cite{Grimvall1981}.
The electron-phonon interaction yields a renormalization of the electronic quasiparticles near the FS, modifying their effective mass and their lifetime, which ultimately gives rise to observable macroscopic phenomena such as the temperature dependent resistivity \cite{Ziman1960},
or even conventional superconductivity \cite{Schrieffer1964, AllenMitrovic1983}.

The ubiquitous presence of the electron-phonon interaction has led to a persistent effort to model this physical process ever since the early days of the quantum theory of solids.
However, practical \textit{ab initio} calculations with the ability to accurately predict complex materials properties related to the electron-phonon interaction have been possible only recently \cite{GiustinoRMP2016}.
The main obstacle to overcome has been the ability to compute, at a reasonable cost, the electron-phonon matrix elements on dense meshes sampling the Brillouin Zone (BZ), which is necessary to capture the fine details of the FS anisotropy.
Several efficient numerical techniques have been developed during the last years for this purpose \cite{BaroniRMP2001,GiustinoPRB2007,EigurenPRB2008},
which have boosted tremendously the accuracy of theoretical studies on electron-phonon driven phenomena, such as
temperature-dependent charge transport \cite{MustafaPRB2016,PoncePRB2018},
non-adiabatic corrections to phonon dispersions \cite{CalandraPRB2010,NovkoPRB2018,GoiricelayaPRB2020},
quasiparticle renormalization signatures in angle resolved photoemission spectra \cite{EigurenPRL2003,VerdiNCM2017,GoiricelayaCMP2019},
or gap anisotropy in phonon-mediated superconductors \cite{MarginePRB2013}.

Nevertheless, these studies have also shown explicitly that extremely fine samplings of the BZ are necessary to obtain converged results,
requiring more than $10^{5}$ points in the reciprocal space in typical cases \cite{GiustinoPRB2007}.
Apart from the obvious issues related to computer memory demands,
the amount of information to be handled on the Fermi surface makes the data analysis of anisotropic quantities certainly difficult,
and relegates the comparison between calculations performed with different meshes to a qualitative level.
More importantly, for electron-phonon problems such as the Eliashberg equations of superconductivity \cite{AllenMitrovic1983, MarginePRB2013},
in which integral equations have to be solved self-consistently on the FS,
the computational workload gets exceedingly high.
This has made the high-throughput calculations of superconducting properties a challenging task up to date.
Thus, developing a method to effectively treat the anisotropy of the electron-phonon interaction while keeping full accuracy seems very appealing.

Almost half a century ago, Allen proposed a procedure by which scalar quantities defined on the FS could be transformed into a new basis set composed of polynomials of electron velocities orthogonalized on the FS,
which he called Fermi-surface harmonics (FSH) \cite{AllenPRB1975}.
He further anticipated that, if the expansion of anisotropic quantities on FSHs was rapidly convergent, integral problems like the Eliashberg equations could be solved in a particularly simple and efficient way.
Despite the interest that the potential of the method sparked in the community,
it has only been applied after imposing further approximations in the anisotropy of the electron-phonon interaction \cite{Butler1976},
or only very recently for relatively simple systems in scarce occasions \cite{HeidPRL2008,XuPRL2014}.
Among the possible reasons behind the lack of systematic applicability of the method are the difficulty in the construction of the basis set,
which involves several semi-analytic steps and requires a different procedure for each crystal structure,
and the fact that the completeness of the basis set cannot be guaranteed for general surfaces.

An alternative definition of the FSH basis set was put forward by some of the authors in Ref.~\cite{EigurenNJP2014}, which overcome the limitations of Allen's proposal.
In this novel approach, the orthonormal basis functions, called Helmholtz Fermi-surface harmonics (HFSH), are obtained by a purely numerical procedure as eigenfunctions of the Laplace-Beltrami operator on a triangularly tessellated Fermi surface,
allowing for a systematic construction of the basis set on any FS topology.
However, the crystal symmetries were incorporated only approximately in the triangulated FS
---and as a result in the properties of the basis set---,
limiting the potential of the method in the compression of physical anisotropic quantities.

In this paper, we improve on Ref.~\cite{EigurenNJP2014} by incorporating the symmetries of the crystal in the HFSH basis set.
We propose a numerical procedure to obtain a fully-symmetric triangulated FS, which is general and applicable to any crystal structure.
The main outcome of the upgrade is the ability to detect functions within the HFSH set that are invariant under all the symmetry operations of the crystal.
As any scalar physical quantity defined on the FS must also follow this symmetry restriction, its expansion on the HFSH basis set will have finite coefficients only in the fully-symmetric subset.
This implies an extra reduction of about an order of magnitude in the compression of anisotropic FS quantities with respect to Ref.~\cite{EigurenNJP2014}.
We describe the procedure using FCC-Cu as an example, and demonstrate its potential in the electron-phonon problem by showing that the mass-enchancement parameter of the anisotropic superconductors HEX-MgB2 and BCC-YH6 can be represented to high accuracy by a handful of coefficients.

The rest of the paper is structured as follows. In Sec.II, we describe the details of the implementation to obtain a fully symmetric triangulated Fermi surface. In Sec.III, we analyze the effects of the symmetries of the triangulated mesh in the HFSH basis set. Using FCC-Cu as an example, we detect the fully symmetric HFSH subset, and confirm that only coefficients of this subset contribute to the expansion of any symmetric quantity defined on the FS. In Sec.IV, we apply the method to two distinct phonon-mediated superconductors with different crystal symmetries and FS topologies, namely HEX-MgB$_2$ and BCC-YH$_6$, demonstrating the general validity and the potential of the method.


\section{Fully symmetric triangulated Fermi surface} \label{sec:trisurf}

 \begin{figure*}[ht]
 \includegraphics[width=2.0\columnwidth]{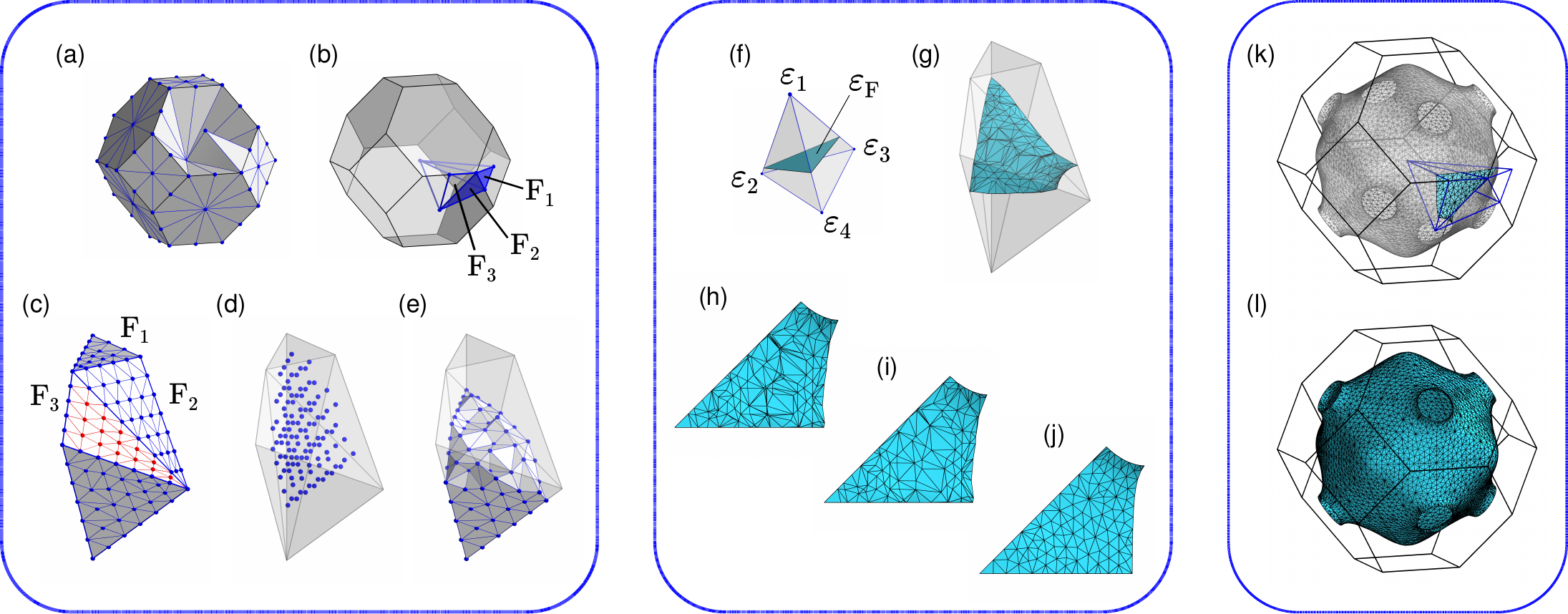}
 \caption{Numerical procedure to obtain a fully symmetric triangulated Fermi surface.
          (a) Star tetrahedral tessellation of the full BZ, from which the irreducible BZ volume can be detected (b). Irreducible faces at the IBZ boundary are highlighted in blue.
          (c),(d),(e) Fine tetrahedral tessellation of the IBZ. The IBZ boundary faces are first triangulated (c), extra Steiner points are added within the IBZ volume (d), and a Delaunay tetrahedralization is performed constrained by the triangular facets at the boundary (e).
          The linear tetrahedron method is applied in the fine tetrahedra (f), from which a triangulated irreducible Fermi surface is obtained (g).
          (h),(i),(j) Mesh-refinement techniques are applied to the IFS, resulting in a high-quality triangulated mesh.
          (k),(l) In a last step, all the symmetry operations are applied to the IFS (k), obtaining a fully symmetric triangulated FS (l).
 \label{fig:Fig1}}
 \end{figure*}

In this section we describe a numerical procedure to obtain from first principles a triangular tessellation of the Fermi surface of any metal, which fulfills all the point group symmetries of its crystal structure.

In principle, a robust method for accomplishing such a task is the linear tetrahedron method \cite{BlochlPRB1994}.
In its original formulation, the tetrahedral tessellation of the BZ is performed in crystal coordinates, where a $\mathbf{k}$-point grid translates into cubes which can be trivially decomposed in six tetrahedra.
This approach was used in Ref.~\cite{EigurenNJP2014}.
However, when analyzing in detail the resulting triangulated isosurface, one finds that the symmetries of the crystal are not incorporated in the FS.
In other words, the triangulated FS obtained is not invariant under all the symmetry operations of the crystal ---as it should---, due to the broken symmetries introduced by the initial tetrahedral tessellation of the BZ.

This led us to modify the original method, and to find an irreducible isosurface in a previously detected irreducible BZ in Cartesian coordinates.
As we will see in the next sections, this variation provides an elegant and effective way of obtaining a fully symmetric triangulated Fermi surface, but also poses some technical difficulties, which are nevertheless overcome by our procedure.

We describe our approach following several steps.
First, the irreducible volume of the BZ (IBZ) is identified.
Then a tetrahedral tessellation of the IBZ is generated, from which a triangulated irreducible Fermi surface (IFS) is obtained using the linear tetrahedron method.
As an optional intermediate step, different mesh improvement techniques are proposed and implemented in order to increase the quality of the triangular mesh.
Finally, the IFS is rotated using all the symmetries of the crystal, resulting in a high-quality and fully symmetric triangulated Fermi surface.

\subsection{Detection of the irreducible wedge of the Brillouin zone}

In any crystal system, an irreducible wedge of the Brillouin zone exists from which the full BZ can be recovered by applying all the symmetry operations that the crystal possesses.
The first task in our procedure will be to identify such an irreducible volume of the Brillouin zone, for any system crystallizing in a given space group.

Geometrically speaking, the BZ is a polyhedron composed of polygonal faces joined by edges.
We first make the observation that, apart from the $\Gamma$ point that lies in the center of the BZ, the high symmetry points of the 14 types of Bravais lattices always lie either in the center of a face, or in the corner or the middle-point of an edge \cite{AroyoAC2014}.
As an illustrative example, we show in Fig.~\ref{fig:Fig1}(a) the BZ of the FCC lattice (space group $\mathrm{Fm}\overline{3}\mathrm{m}$), in which all the corners, the centers of the faces and the middle-points of the edges have been highlighted with blue dots.

Joining each of the points in the edges with the points at their nearest corners, and these two in turn with the points at the center of their corresponding face, we can create a triangular tessellation of the polygonal faces of the BZ.
Moreover, joining all of these points with the $\Gamma$ point in the center of the BZ, we can obtain a star tetrahedral tessellation of the whole BZ volume.
Given that, by definition, all the non-equivalent high symmetry points have to be included in the irreducible wedge of the BZ,
we can represent the IBZ as the sum of several of these tetrahedra.
We show in Fig.~\ref{fig:Fig1}(a) the coarse tetrahedral tessellation of the FCC BZ volume obtained in this way, in which some tetrahedra have been removed for ease of visualization.

In order to determine which are the irreducible tetrahedra that we have to include in order to form the IBZ of a given system, we need to know the particular symmetry operations belonging to its space group.
The procedure is similar to the one used to detect the irreducible number of ${\bf k}$-points within a regular grid that can form a full mesh in the BZ by applying all the symmetry operations of a particular system.
In a first step, one selects an arbitrary tetrahedron and applies all the symmetry operations allowed by the point group.
In this way, we detect the volume of the BZ connected by symmetry to the initially selected tetrahedron.
We repeat this operation for all the tetrahedra in the initial tessellation of the BZ volume, constructing in this way the irreducible volume of the BZ.
As an example, the resulting IBZ volume for the FCC lattice, which is composed by three tetrahedra, is shown in Fig.~\ref{fig:Fig1}(b).

As a result of the translational invariance of crystals,
at the BZ boundary the Fermi surface possesses extra symmetries beyond the point group.
In this respect, we have to check for further reduction of the irreducible wedge at the BZ boundary.
For this purpose, we repeat a similar procedure as the one described above, but only for the triangular facets on the boundary of the IBZ volume.
We now apply $\mathcal{S}+\mathbf{G}$ operations, where $\mathcal{S}$ is a symmetry rotation and $\mathbf{G}$ is a reciprocal lattice vector, and check if any of the facets can be recovered from an irreducible subgroup.
Following with the FCC example, we find that one of the three triangular facets ($\mathrm{F}_{3}$) can be recovered in this way from its neighbor facet ($\mathrm{F}_{2}$). The irreducible facets of the IBZ ($\mathrm{F}_{1}$ and $\mathrm{F}_{2}$) are highlighted in blue in Fig.~\ref{fig:Fig1}(b).

\subsection{Tetrahedral tessellation of the irreducible wedge of the Brillouin zone}

The next step in our procedure will be to obtain a fine tetrahedral tessellation of the irreducible wedge of the Brillouin zone identified in the previous section.
The tetrahedral tessellation of a general polyhedron defining the IBZ in Cartesian coordinates is not trivial, and the  $\mathcal{S}+\mathbf{G}$ symmetries mentioned in the previous section forces us to proceed with care.
We describe the scheme we have implemented for this purpose in the following.

The first task will be to triangulate the faces of the IBZ volume in such a way that all the $\mathcal{S}+\mathbf{G}$ symmetries are fulfilled.
To this end, in a first step we triangulate the irreducible facets that are related to the non-irreducible ones by symmetry ($\mathrm{F}_{2}$ in Fig.~\ref{fig:Fig1}(c)).
In a second step, we obtain the triangulation of the non-irreducible facets by applying the corresponding symmetry operations ($\mathrm{F}_{3}$ in Fig.~\ref{fig:Fig1}(c)).
In a third step, we triangulate all the rest of the irreducible facets, considering the constraints imposed by the nodes already present in the facet-joining edges.

For the triangulation of each facet, we first distribute nodes throughout the facet-plane. This distribution is done in such a way that the projection of a given mesh of points in the reciprocal-lattice vectors $\left\lbrace n_{k_{1}},n_{k_{2}},n_{k_{3}}\right\rbrace$ onto the facet-plane is approximately matched, setting the condition that the nodes on each edge are regularly spaced.
Then a constrained Delaunay triangulation is constructed from these nodes using the {\sc Triangle} code \cite{ShewchukSpringer1996}, in which the edges of the facet are maintained. As an example, the triangulation obtained in such a way for the boundary of the FCC IBZ volume is shown in Fig.\ref{fig:Fig1}(c), where the facet obtained by symmetry is highlighted in red.

Next, as shown in Fig.~\ref{fig:Fig1}(d), we populate the IBZ volume with a set of regularly spaced points, selected from the points of the $\left\lbrace n_{k_{1}},n_{k_{2}},n_{k_{3}}\right\rbrace$ mesh that fall within this volume.

Finally, a constrained Delaunay tetrahedralization is constructed using the {\sc TetGen} code \cite{SiACM2015}, in which the boundary triangulation is maintained and the volume-nodes are added as Steiner points. The resulting tetrahedral tessellation of the FCC IBZ example is shown in Fig.~\ref{fig:Fig1}(e), in which some of the tetrahedra on the upper part have been removed for ease of visualization.

\subsection{Linear tetrahedron method and triangle mesh refinement}

Filling the IBZ volume with tetrahedra, as described above, allows us to apply the linear tetrahedron method \cite{BlochlPRB1994} in order to obtain a numerical representation of the irreducible Fermi surface in terms of a triangle mesh.

Each tetrahedron marks four points in the reciprocal space in which the energies have to be computed, as represented schematically in Fig.~\ref{fig:Fig1}(f) by $\left( \varepsilon_{1},\varepsilon_{2},\varepsilon_{3},\varepsilon_{4}\right)$.
Then we check if the energy corresponding to the isosurface lies within the values at the corners of the tetrahedron.
In the affirmative case, a linear interpolation among the values at the corners gives an approximation to the points in which the isosurface crosses the tetrahedral edges.
Depending on the number of edges that the isosurface crosses, one or two triangles can be formed inside the tetrahedron, as discussed in detail, for example, in Ref.\cite{EigurenNJP2014}.
The simplest case is shown in Fig.~\ref{fig:Fig1}(f), in which the isosurface (denoted as $\varepsilon_{\mathrm{F}}$) crosses three of the tetrahedral edges, directly forming a triangle inside.
All the triangles constructed in this way form a two-dimensional triangle mesh, representing numerically the Fermi surface within the IBZ.

As an illustrative example, we show in Fig.~\ref{fig:Fig1}(g) the isosurface obtained for FCC-Cu from the tetrahedral tessellation of the IBZ shown in Fig.~\ref{fig:Fig1}(e)
\footnote{
The ground state calculations for FCC-Cu are performed within the generalized gradient approximation of density functional theory \cite{PerdewPRL1996} on a coarse $12\times12\times12$ ${\bf k}$-point grid using the {\sc Quantum ESPRESSO} package \cite{QE2017},
and the energies at the tetrahedral vertices are obtained by means of Wannier interpolation \cite{MarzariPRB1997,SouzaPRB2001,PizziJPCM2020}.
} .
A clearer view of the triangle mesh formed in this example is shown in Fig.~\ref{fig:Fig1}(h).

As it can be noted from this figure, even if a good initial tetrahedral tessellation is provided,
the resulting triangular mesh may be of a low quality, meaning that the isosurface may present an inhomogeneous density of vertices which will most likely form a set of triangles with a poor aspect ratio.
Although not strictly necessary, it is highly desirable to incorporate procedures to improve the quality of the mesh,
possibly eliminating redundant and poor-quality triangles.
We have implemented two different mesh refinement techniques, namely the mesh-simplification and the vertex-relaxation procedures \cite{botschCRC2010}.
Special care has been taken with the vertices at the BZ boundary, so that the borders of the irreducible Fermi surface are preserved, and the $\mathcal{S}+\mathbf{G}$ symmetries are maintained after the refinement process.

In the mesh-simplification procedure, triangles with a poor shape-quality are detected first, following the criteria that one of their edges is much shorter than the perimeter of the triangle, up to a given threshold value.
This short edge is collapsed, so that one vertex and one triangle are removed from the mesh, though maintaining the original topology.
This procedure is repeated iteratively until all poor shape-quality triangles are eliminated.
The simplified mesh obtained after this procedure in the FCC-Cu example is shown in Fig.~\ref{fig:Fig1}(i).

The so-called vertex-relaxation procedure consists of two steps.
First, a tangential relaxation of the vertices is performed.
Each vertex is moved from its position seeking a homogeneous distance with respect to all of its neighbor vertices.
However, this movement is constrained to the tangential plane of the vertex, defined by its velocity vector, $\mathbf{v}_{n\mathbf{k}}=\nabla \varepsilon_{n\mathbf{k}} / \hbar$.
Note that this vector for a $\mathbf{k}$-point at the Fermi surface is, by definition, the normal vector of the Fermi surface at this point.
The Fermi velocities $\mathbf{v}_{n\mathbf{k}}$ at the triangular vertices are computed efficiently by means of the Wannier interpolation method \cite{MarzariPRB1997,SouzaPRB2001,PizziJPCM2020}.
This procedure is repeated iteratively for all the vertices in the mesh, resulting in a homogeneous distribution of triangles with similar areas.
Finally, the vertices are relaxed along their normal vector.
This additional step compensates the error introduced by the linear interpolation in the regular linear tetrahedron method, so that the final relaxed vertices are located at $\varepsilon_{F}$ to a great accuracy \cite{EigurenNJP2014}.

The final refined mesh for the FCC-Cu example is shown in Fig.~\ref{fig:Fig1}(j), where the improvement in the quality of the mesh is clearly appreciated.
These mesh refinement techniques translate into a considerable accuracy and efficiency gain in the computation of Fermi surface integrals.

\subsection{Rotation to a fully symmetric Fermi surface}

The very last step in our procedure consists of applying all the symmetry operations to the irreducible Fermi surface described in the previous sections,
in order to obtain a fully symmetric Fermi surface mesh,
which is invariant under all the symmetry operations of the crystal up to numerical precision.

The irreducible part of the Fermi surface of the FCC-Cu example is shown within the full BZ in Fig.~\ref{fig:Fig1}(k).
The complete Fermi surface mesh obtained by rotation of the irreducible part is shown in Fig.~\ref{fig:Fig1}(l).
As can be appreciated in the figure, our procedure provides a high-quality triangulated Fermi surface, which fulfills all the symmetries of the crystal.
We note that working on the IBZ as a prior step, and making use of efficient computational geometry packages \cite{ShewchukSpringer1996,SiACM2015} and the Wannier interpolation method \cite{MarzariPRB1997,SouzaPRB2001,PizziJPCM2020}, turns the computational cost of constructing the fully symmetric triangular mesh minimal as compared, for example, to a typical ground state calculation.

As a last remark, we mention that even though the linear tetrahedron method has been already used in other works to obtain a triangulated Fermi surface \cite{KawamuraCPC2019,MustafaPRB2016,RittwegerJPCM2017}, to the best of our knowledge the symmetries of the FS have not been incorporated exactly in the triangulated mesh in these works, as it is done in our procedure.
In the next sections we will show the importance of this aspect in the construction of the Helmholtz Fermi-surface harmonics basis set,
and in its application to compress anisotropic quantities on the Fermi surface into a few coefficients.


\section{Symmetries on the Helmholtz Fermi-surface harmonics basis set} \label{sec:symfsh}

In this section, we show how the fully symmetric triangulated Fermi surface obtained by the procedure described in Sec.~\ref{sec:trisurf} provides a direct way to incorporate the symmetries of the crystal in the HFSH basis set.

For completeness, we first review the main aspects of the method proposed in Ref.~\cite{EigurenNJP2014} to obtain the basis set.
The HFSHs are defined as the eigenmodes of a velocity-weighted Laplace-Beltrami operator on the curved Fermi surface,
\begin{equation} \label{eq:lapl-belt}
  v({\bf k}) \nabla^{2}_{\bf k} \Phi_{L}({\bf k}) + \omega_{L} \Phi_{L}({\bf k}) = 0 ,
\end{equation}
where $\omega_{L}$ are the eigenvalues associated with the HFSH set functions $\left\lbrace \Phi_{L}({\bf k}) \right\rbrace$, which obey the following orthogonality condition,
\begin{equation} \label{eq:orth_cond}
  \int \frac{d^{2}s_{\bf k}}{v({\bf k})} \Phi_{L'}({\bf k}) \Phi_{L}({\bf k}) = \delta_{L',L} \int \frac{d^{2}s_{\bf k}}{v({\bf k})} ~.
\end{equation}

The triangular tessellation of the Fermi surface allows for a numerical solution of a discretized version of Eq.~\ref{eq:lapl-belt}, which is transformed into a generalized sparse eigenvalue problem:
\begin{equation} \label{eq:discr-lapl-belt}
  \frac{v({\bf k}_{i})}{S_{i}} \sum_{j} \Omega_{i,j} \Phi_{L}({\bf k}_{j}) = \omega_{L} \Phi_{L}({\bf k}_{i})~.
\end{equation}
In this expression, $i$ and $j$ are indices for vertices on the triangulated mesh, ${\bf k}_{i}$ represents the coordinates of a vertex $i$ in the reciprocal space, and $S_{i}$ its control area, defined as the sum of $\frac{1}{3}$ of its neighboring triangle areas.
The discretized Laplace-Beltrami operator $\Omega_{i,j}$ takes the form
\begin{equation} \label{eq:discr-lapl-omegaij}
  \Omega_{i,j} =
  \begin{cases}
      -\frac{1}{2} \left[ \cot(\alpha_{i,j}) + \cot(\beta_{i,j}) \right]  & i\ne j\\
      \sum_{i\ne j} \Omega_{i,j}                                          & i=j
  \end{cases}
  ~,
\end{equation}
where $\alpha_{i,j}$ and $\beta_{i,j}$ are the two opposite angles of the triangles sharing the edge joining the vertices $i$ and $j$.

By virtue of its completeness, we can efficiently represent any anisotropic function $F({\bf k}_i)$ defined on the triangulated Fermi surface by performing an expansion in the HFSH basis set,
\begin{equation} \label{eq:fsh_expansion}
  F({\bf k}_i) = \sum_{L} c_{L}(F) \Phi_{L}({\bf k}_{i}) ~,
\end{equation}
where the expansion coefficients are defined by the following FS integrals:
\begin{equation} \label{eq:fsh_coef}
  c_{L}(F) \equiv \frac{\int_{S_{F}} \frac{d^{2}s_{\bf k}}{v({\bf k})} \Phi_{L}({\bf k}) F({\bf k}) } { \int_{S_{F}} \frac{d^{2}s_{\bf k}}{v({\bf k})} }
  \approx
  \frac{ \sum_{i} \frac{S_{i}}{v({\bf k}_i)} \Phi_{L}({\bf k}_i) F({\bf k}_i) }{ \sum_{i} \frac{S_{i}}{v({\bf k}_i)} } ~.
\end{equation}

We refer the reader to Ref.~\cite{EigurenNJP2014} for further details and properties of the HFSH basis set.

\subsection{Degenerate subspaces}

Clearly, the symmetries of the surface on which Eq.~\ref{eq:lapl-belt} is defined translate into symmetric properties of the HFSH basis set.
Given that the symmetries of the surface are exactly maintained in the triangulated mesh,
these properties will be preserved in the discretized form of Eq.~\ref{eq:discr-lapl-belt}.
The fulfillment of this requirement is guaranteed if the mesh is constructed following the procedure described in Sec.~\ref{sec:trisurf}.
We note that any symmetric mesh obtained by any other alternative procedure is also valid for the conclusions drawn in the rest of the paper.

A straight consequence of retaining the symmetries on the surface appears in the degeneracies of the energy levels $\omega_{L}$.
For instance, in a perfect sphere, the full rotational symmetry enforces the threefold and fivefold degeneracies in the $p$ and $d$ spherical harmonics, respectively.
Even though the full rotational symmetry of the sphere is broken in a realistic Fermi surface due to the crystal field, the possible discrete rotational symmetries of the crystal may enforce subspaces within the HFSH basis set which are exactly degenerate.

\begin{figure}[t]
 \includegraphics[width=0.9\columnwidth]{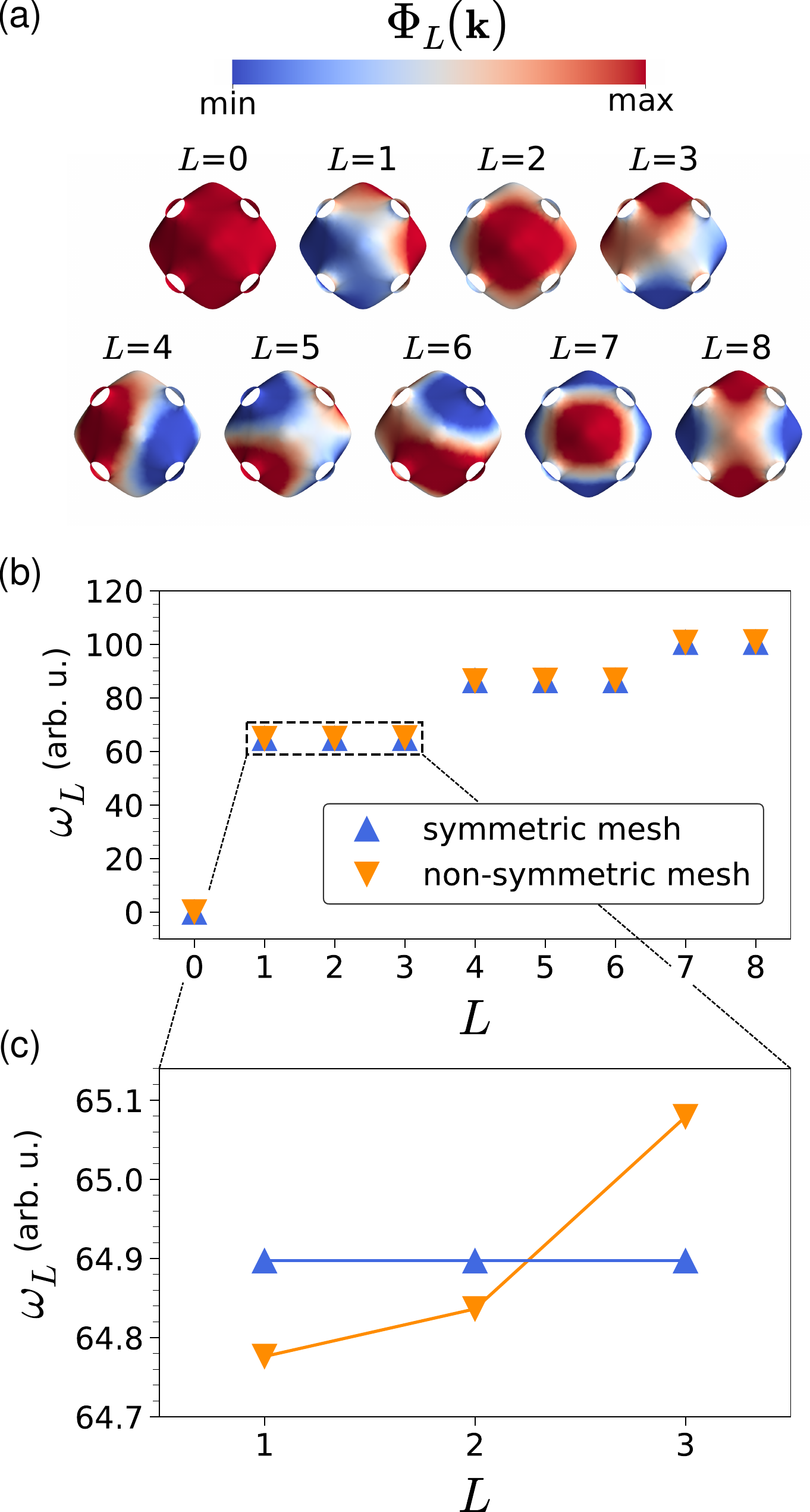}
 \caption{Degeneracies in the HFSH eigenmodes.
          (a) First nine HFSH basis functions for FCC-Cu, and (b) their corresponding eigenvalues.
          The eigenvalues obtained using the fully symmetric mesh are shown in blue, and the eigenvalues obtained in Ref.~\cite{EigurenNJP2014} are shown in orange for comparison.
          (c) Zoom on the first non-zero eigenvalues, highlighting the numerical accuracy of the degeneracy in the symmetric mesh.
 \label{fig:Fig2}}
\end{figure}

Continuing with the FCC-Cu example, we show in Fig.~\ref{fig:Fig2}(a) the first nine HFSH basis functions $\Phi_{L}({\bf k})$, obtained as solutions of Eq.~\ref{eq:discr-lapl-belt} on the symmetric mesh produced in Sec.~\ref{sec:trisurf}.
The corresponding eigenvalues $\omega_{L}$ are shown in Fig.~\ref{fig:Fig2}(b), compared with the eigenvalues obtained on a mesh in which the crystal symmetries are not explicitly enforced, as in Ref~\cite{EigurenNJP2014}.
As discussed in Ref.~\cite{EigurenNJP2014}, the threefold degeneracy in the $p$-like harmonics is maintained, but the energies of the $d$-like states are split into two subspaces of threefold and twofold degeneracies.

However, a closer look reveals that these degeneracies are fulfilled only approximately on the non-symmetric mesh, as shown in Fig.~\ref{fig:Fig2}(c) for the energies of the $p$-like harmonics.
As we can see in this figure, not incorporating the symmetries exactly on the mesh can introduce errors of $\sim 0.15\%$ in the energies.
In contrast, when using the exactly symmetric mesh we obtain equal energies up to numerical accuracy, with relative differences of the order of $\sim 10^{-10}$ in this particular example.
Similar results are obtained for all the degenerate subspaces of the full HFSH basis set.

\subsection{Fully symmetric HFSHs} \label{sec:fullsymhfsh}

\begin{figure}[b]
 \includegraphics[width=1.0\columnwidth]{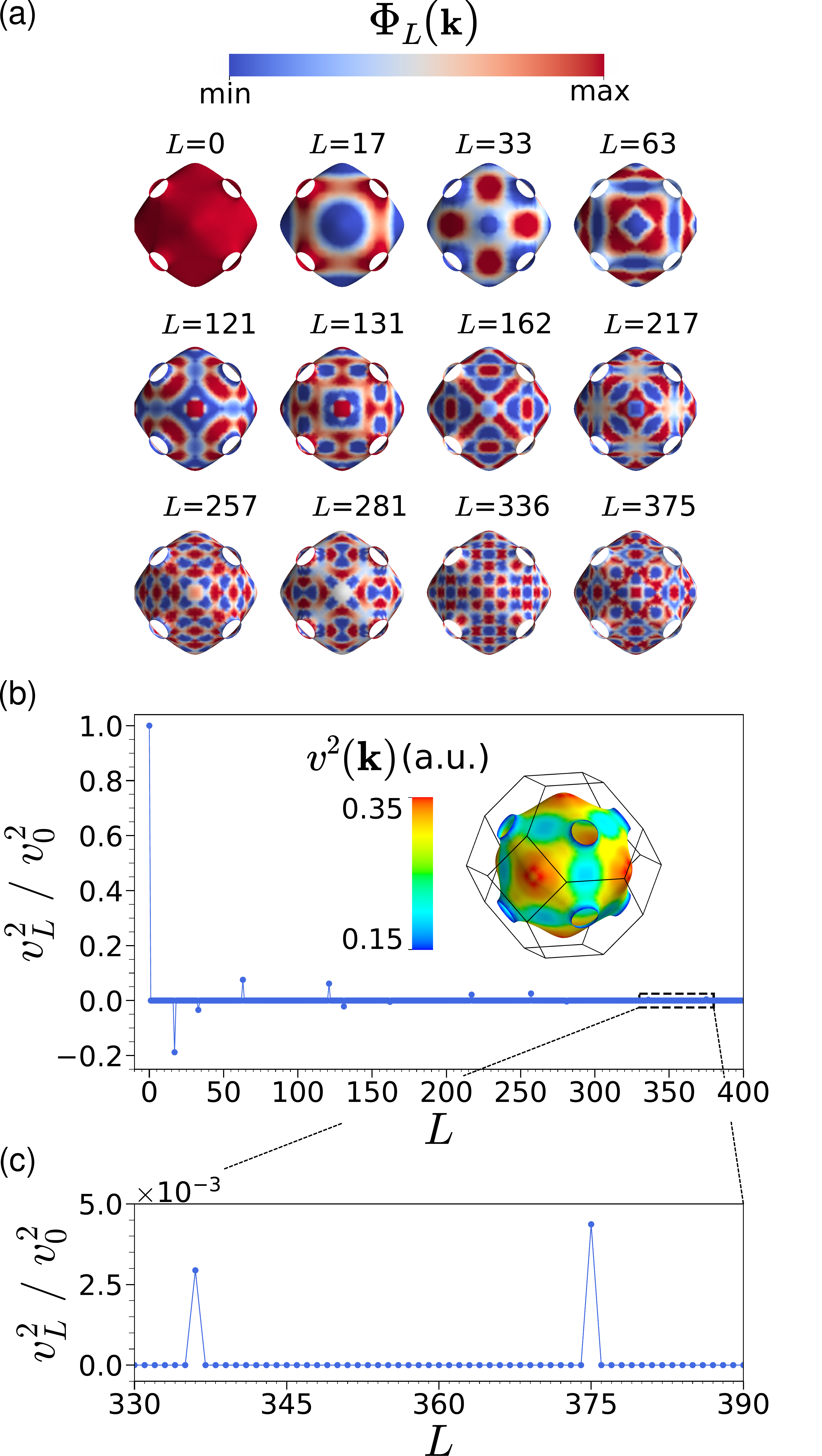}
 \caption{Fully symmetric HFSHs.
          (a) First 12 fully symmetric HFSH basis functions for FCC-Cu, together with their index in the full HFSH basis set.
          (b) First 400 expansion coefficients in the HFSH basis set for the squared modulus of the electron velocity, relative to the $L=0$ coefficient. The magnitude and anisotropy of $v^{2}({\bf k})$ over the FS is shown in the inset in atomic units.
          Only the fully symmetric HFSH functions shown in (a) give finite contributions, as highlighted in
          (c) where a zoom on the last two finite coefficients is shown.
 \label{fig:Fig3}}
\end{figure}

As the triangular mesh in which Eq.~\ref{eq:discr-lapl-belt} is solved and in which the $\Phi_{L}({\bf k}_{i})$ functions are defined is exactly symmetric, we can identify numerically those functions within the HFSH basis set that are invariant under all the symmetry operations of the crystal.
We will name this subset the \emph{fully symmetric} HFSHs, and label them with the symbol $\tilde{L}$.
Formally, they are identified as the functions within the HFSH basis set that satisfy the following condition:
\begin{equation} \label{eq:fulsym_hfsh}
  \Phi_{\tilde{L}}(\mathcal{S}_{n}{\bf k}_{i}) = \Phi_{\tilde{L}}({\bf k}_{i}) ~,
\end{equation}
for all $n$, where $\mathcal{S}_{n}$ is a symmetry operation of the crystal.
As an example, in the FCC-Cu case, out of the first 400 HFSH functions only 12 satisfy Eq.~(\ref{eq:fulsym_hfsh}) and are shown in Fig.~\ref{fig:Fig3}(a).

The fact that most of the physical properties defined on the Fermi surface are invariant under all the symmetry operations of the crystal imposes severe restrictions on their expansion in the HFSH basis set.
As can be directly deduced from Eq.~\ref{eq:fsh_expansion}, if a given function $F({\bf k}_i)$ is fully symmetric on the FS, only those coefficients corresponding to the fully symmetric HFSHs can give a finite contribution in the expansion.

We now demonstrate that this restriction is satisfied in our implementation up to numerical precision. As an illustrative example, we consider the squared of the Fermi velocity, $F({\bf k})=v^{2}({\bf k})$, clearly a fully symmetric function [see inset of Fig.~\ref{fig:Fig3}(b)].
We show in Fig.~\ref{fig:Fig3}(b) the first 400 coefficients of the expansion of this function in the HFSH basis set [see Eq.~\ref{eq:fsh_coef}], relative to the value of the first coefficient, i.e. the FS average $v^{2}_{0}\equiv\langle v^{2}({\bf k}) \rangle_{\mathrm{FS}}$.

As appreciated from this figure, the values of the expansion coefficients decrease rapidly for larger HFSH indices.
This trend goes in line with the fact that HFSHs with higher energies oscillate more intensely, and therefore only add finer details to the anisotropy of the expanded function.
The more isotropic is the quantity to be transformed, the less are the coefficients needed for a faithful representation of its anisotropy.
In the extreme case of a constant function, only the first coefficient will be finite.

Most importantly, we see that only those coefficients corresponding to the fully symmetric HFSHs shown in Fig.~\ref{fig:Fig3}(a) have a finite value, the rest being all strictly zero up to numerical precision.
We show this more clearly in Fig.~\ref{fig:Fig3}(c), which zooms into the last two finite coefficients of Fig.~\ref{fig:Fig3}(b).
The magnitude of these coefficients is only $\sim 0.5\%$ of the average value, showing that we can achieve this accuracy in the representation of the anisotropic function $v^{2}({\bf k})$ of this example by using only 12 coefficients.

All in all, the use of a fully symmetric Fermi surface for the construction of the HFSH basis set, and the identification of the fully symmetric HFSH subset, allows us to obtain an extra reduction of at least one order of magnitude in the computational workload to describe anisotropic quantitites on the FS with respect to Ref.~\cite{EigurenNJP2014}.
Note that this method already introduced a saving factor of approximately two orders of magnitude with respect to the conventional ${\bf k}$-space representation.
In the next section we apply this methodology to the electron-phonon problem, where the ${\bf k}$-space representation of the related quantities generates a major bottleneck in the computation of prominent properties of metals, such as superconductivity.

\section{Electron-phonon anisotropy in the HFSH representation} \label{sec:elphfsh}

\begin{figure*}[ht]
  \includegraphics[width=2.0\columnwidth]{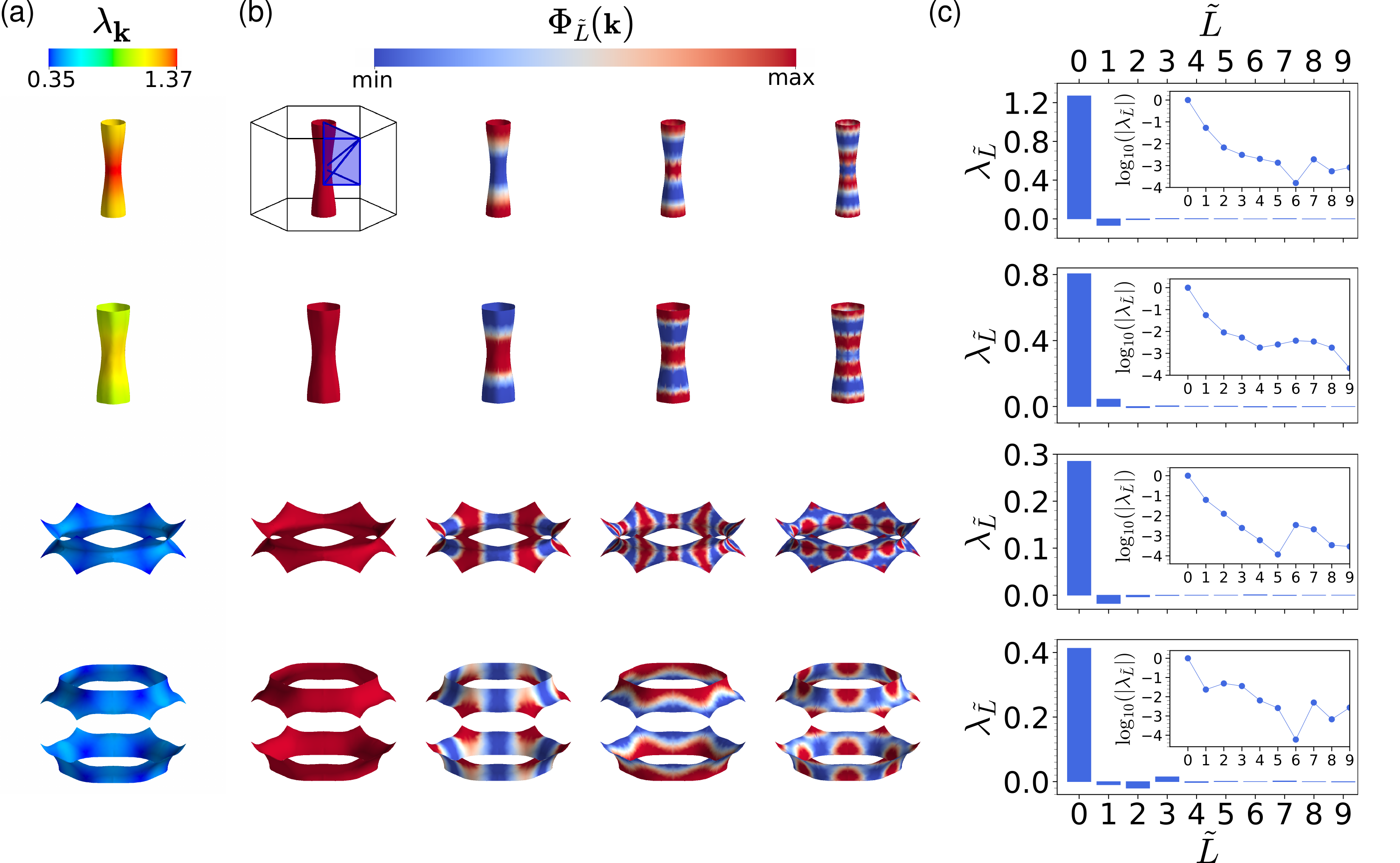}
  \caption{(a) Anisotropic electron-phonon mass-enhancement parameter $\lambda_{{\bf k}}$ on the Fermi surface of MgB$_2$, separated in the four different FS sheets.
           (b) First four fully symmetric HFSH basis functions for each FS sheet. The hexagonal BZ and the corresponding IBZ is shown in the top left corner.
           (c) First ten expansion coefficients of $\lambda_{{\bf k}}$ for each FS sheet in the fully symmetric HFSH subset.
           The inset shows the same result with a logarithmic scale on the $y$-axis.
\label{fig:Fig4}}
\end{figure*}

Given that the energy scale of phonons ($\sim$meV) is roughly three orders of magnitude smaller than the energy scale of the electrons ($\sim$eV), the electron-phonon scattering events are, to a good approximation for most metals, limited to the Fermi surface.
Nonetheless, the variation of the electronic wave functions and velocities on the FS is usually sizable,
and so is the variation of the phonon frequencies and change of potential for the different momentum vectors joining the electron states on the FS.
This implies that, for an accurate description of electron-phonon problems, the anisotropy of the matrix elements and the related quantities on the FS have to be accurately taken into account,
which poses a major computational bottleneck in practical calculations.
In this section we will show that the HFSH representation presented in Sec.~\ref{sec:symfsh} provides an elegant and extremely efficient solution to this difficulty.

As a representative anisotropic quantity related to the electron-phonon problem, we consider the momentum-dependent mass enhancement parameter for electron states at the FS,
\begin{equation}\label{eq:lambda_k}
    \lambda_{n{\bf k}} =
    \frac{2}{\Omega_{\textrm{BZ}}} \sum_{m\nu} \int_{S_{F}} \frac{d^{2}s_{{\bf k}'}}{v({\bf k}')} ~ \frac{|g^{\nu}_{mn}({\bf k},{\bf k}')|^{2}}{\omega^{\nu}_{{\bf k}'-{\bf k}}} ~,
\end{equation}
where $\Omega_{\textrm{BZ}}$ is the BZ volume, $n$ and $m$ represent electron band indices,
$\omega^{\nu}_{{\bf k}'-{\bf k}}$ is the frequency of the phonon mode $\nu$ at momenta ${\bf q}\equiv{\bf k}'-{\bf k}$,
and $g^{\nu}_{mn}({\bf k},{\bf k}')$ is the electron-phonon matrix element for the scattering of an electron $n{\bf k}$ to $m{\bf k}'$ via emission/absorption of a phonon  $\nu{\bf q}~.$
The $\lambda_{n{\bf k}}$ parameter is the most meaningful measure of the quasiparticle renormalization driven by electron-phonon interactions,
directly affecting several transport properties such as the electronic heat capacity or the amplitude of the de Haas-van Alphen oscillations \cite{Grimvall1981}.
Its average over the FS is a central parameter in simplified expressions for the critical temperature of superconductors \cite{McMillanPR1968,AllenDynesPRB1975},
and its two-index and frequency-dependent generalization is crucial in the full Eliashberg theory of superconductivity \cite{AllenMitrovic1983}.
We note, however, that the compression described below is equally applicable to any anisotropic quantity defined on the FS,
such as transport scattering rates $1/\tau_{{\bf k}}$, or the superconducting energy gap $\Delta_{{\bf k}}$.

\subsection{HEX-MgB$_2$}

As a first example we consider the prototypical anisotropic phonon-mediated superconductor MgB$_2$.
The calculation parameters have been chosen with the aim of making the comparison with previous works as direct as possible.
The ground state calculations have been performed with the {\sc Quantum ESPRESSO} package \cite{QE2017} within the local density approximation of density functional theory \cite{PerdewPRB1981} in a $24^{3}$ \textbf{k}-point grid, using norm-conserving pseudopotentials and a kinetic energy cutoff of 60Ry in the plane-wave expansion of valence electronic wave functions.
The lattice parameters have been set to the experimental values of $a=5.832~\mathrm{bohr}$ and $c/a=1.142$ \cite{NagamatsuNAT2001}.
Phonon properties have been computed within density functional perturbation theory \cite{BaroniRMP2001} on a $8^3$ \textbf{q}-point grid.
Electron-phonon matrix elements have been computed on a coarse $(8^{3},8^{3})$ \textbf{k} and \textbf{q}-point grid,
and the Wannier interpolation method \cite{MarzariPRB1997,SouzaPRB2001,PizziJPCM2020,GiustinoPRB2007,EigurenPRB2008} has been used to interpolate the matrix elements to the triangular vertices
\cite{EigurenPRL2008,EigurenPRB2008,EigurenPRB2009,GoiricelayaPRB2018,GoiricelayaCMP2019,GoiricelayaPRB2020}.
As a last remark, we note that the high-quality triangulated Fermi surface as obtained by the method presented in Sec.~\ref{sec:trisurf} allows for an efficient numerical integration of Eq.~(\ref{eq:lambda_k}),
\begin{equation}\label{eq:num_lambda_k}
    \lambda_{n{\bf k}_{i}} \approx
    \frac{2}{\Omega_{\textrm{BZ}}} \sum_{m\nu} \sum_{j} \frac{S_{j}}{v({\bf k}_{j})} ~ \frac{|g^{\nu}_{mn}({\bf k}_{i},{\bf k}_{j})|^{2}}{\omega^{\nu}_{{\bf k}_{j}-{\bf k}_{i}}} ~,
\end{equation}
where only the matrix elements at the ${\bf k}$-points lying on the FS vertices are needed.

We show in Fig.~\ref{fig:Fig4}(a) our results for the anisotropic mass-enhancement parameter of MgB$_2$, in which the four different FS sheets have been separated for clarity.
In agreement with previous works \cite{EigurenPRB2008,ChoiPRB2002,MarginePRB2013}, we find that $\lambda_{n{\bf k}}$ takes considerably large values in the range of $1.00$--$1.37$ on the cylinder-like FS sheets corresponding to the $\sigma$ bands.
In contrast, the FS sheets formed by the $\pi$ bands couple much less efficiently to phonons, resulting in smaller $\lambda_{n{\bf k}}$ values in the range of $0.35$--$0.47$.
Apart from the arrangement of the absolute values of the $\lambda_{n{\bf k}}$ parameter in two main groups,
this figure also shows that its anisotropy within each FS sheet is sizable.

For the Fermi surface averaged mass-enhancement parameter
we obtain $\lambda=0.73$, also in very good agreement with previous calculations \cite{EigurenPRB2008,ChoiPRB2002,MarginePRB2013}.
In particular, we find that our results agree very well with the values presented in Ref.~\cite{MarginePRB2013}, where a systematic convergence test of $\lambda$ with respect to the \textbf{k}-point sampling was performed.
Remarkably, while they showed that $\sim 10^{5}$ points are needed in the three-dimensional BZ to obtain converged results when approximating the FS with a smearing function,
we already obtain convergence in the average value and the distribution of $\lambda_{{\bf k}}$ with $\sim 8\times 10^{3}$ points in the triangulated Fermi surface.

Now we move on to the HFSH representation.
Being a scalar quantity, the $\lambda_{n{\bf k}}$ parameter is invariant under all the symmetry operations of the crystal, as can be appreciated in Fig.~\ref{fig:Fig4}(a).
Therefore, as discussed in Sec.~\ref{sec:fullsymhfsh}, its expansion will only have finite coefficients in the fully symmetric HFSH subset:
\begin{equation} \label{eq:lambda_L}
  \lambda_{n,\tilde{L}} = \frac{\int_{S_{F_{n}}} \frac{d^{2}s_{\bf k}}{v({\bf k})} ~ \Phi_{n,\tilde{L}}({\bf k}) ~ \lambda_{n{\bf k}} } { \int_{S_{F_{n}}} \frac{d^{2}s_{\bf k}}{v({\bf k})} } ~ .
\end{equation}
Note that the HFSH functions for each FS sheet are independent by construction \cite{EigurenNJP2014} ,
and that the integrals are performed over the corresponding sheet $S_{F_{n}}$.

We show in Fig.~\ref{fig:Fig4}(b) the first four fully symmetric HFSH functions for the different FS sheets of MgB$_2$.
The first HFSH function is always the trivial constant solution with eigenvalue $\omega_{\tilde{L}}=0$, and the following ones oscillate more and more rapidly in direct analogy with the normal modes of a vibrating membrane.
The first ten $\lambda_{n,\tilde{L}}$ coefficients given by Eq.~(\ref{eq:lambda_L}) are shown in Fig.~\ref{fig:Fig4}(c).
As it can be anticipated from  Eq.~(\ref{eq:fsh_expansion}) and by looking at the HFSH functions of Fig.~\ref{fig:Fig4}(b), the $\tilde{L}=$~$0$ coefficient gives the average value of $\lambda_{n{\bf k}}$ in each FS sheet, and the subsequent coefficients add finer and finer anisotropic details.
It is therefore reassuring to see that the magnitude of the $\lambda_{n,\tilde{L}}$-s decay very quickly for bigger $\tilde{L}$-s.
What is more remarkable is the rate at which the coefficients decay.
In order to analyze this point further, we plot in the insets of Fig.~\ref{fig:Fig4}(c) the same result but using a logarithmic scale in the $y$ axis,
revealing that the value of the coefficients decay very rapidly.

Indeed, this result demonstrates that the transformation from \textbf{k}-space to the HFSH representation turns out strikingly beneficial,
as all the details of the $\lambda_{n{\bf k}}$ parameter can be compressed with an accuracy of at least $10^{-3}$ in as few coefficients as $10~\lambda_{\tilde{L}}$-s per FS sheet.
In comparison with the $\sim 8000$ triangular vertices needed in \textbf{k}-space, this simplification implies a saving factor of $\sim 2\times 10^{2}$ with no loss of accuracy.

\begin{figure}[hb!]
  \includegraphics[width=1.0\columnwidth]{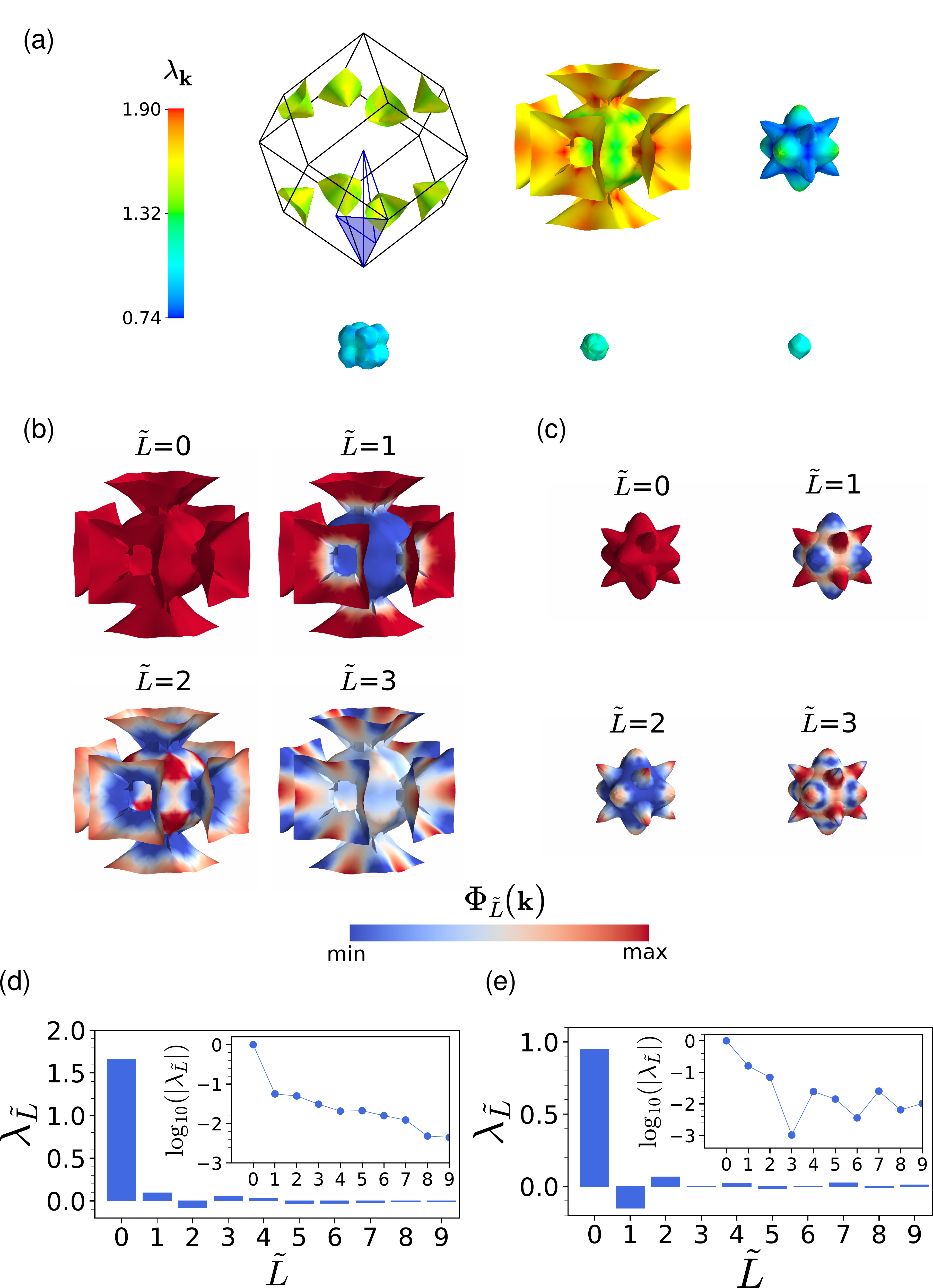}
	\caption{(a)Anisotropic electron-phonon mass-enhancement parameter on
	the six different FS sheets of YH$_6$ at 300GPa.
	The BCC Brillouin zone and the corresponding IBZ is shown in the top left corner.
	First four fully symmetric HFSH basis functions for the (b) second and (c) third FS sheets of BCC-YH6.
	First ten expansion coefficients of $\lambda_{{\bf k}}$ for the (d) second and (e) third FS sheets in the fully symmetric HFSH subset.
        The insets show the same result with a logarithmic scale on the $y$-axis.
\label{fig:Fig5}}
\end{figure}

\subsection{BCC-YH$_6$}

In order to validate that our method is robust and applicable to different systems,
we apply the very same procedure presented in the previous sections to BCC-YH$_6$.

Compressed hydrides have been attracting an enormous interest during the last years,
especially since the prediction and discovery of conventional high-temperature superconductivity in hydrogen sulfide under high pressures \cite{LiJCP2014,DuanSCR2014,DrozdovNAT2015}.
Developing methods to alleviate the computational cost of calculating superconducting properties appears particularly interesting in this research field, in which theoretical predictions of new candidates guide the experimental efforts to find materials with increasingly favorable properties \cite{FloresLivasPHR2020}.
YH$_6$ represents an interesting case within this class of materials,
as the very recent experimental confirmation of superconductivity in this system \cite{TroyanARXIV2019,KongARXIV2019} has revealed a sizable deviation in the measured critical temperature with respect to the theoretical predictions \cite{LiSCR2015,PengPRL2017,HeilPRB2019}.

We have considered YH$_6$ at 300GPa in the BCC structure, whith the lattice parameter reported in Ref.~\cite{HeilPRB2019}.
Ground state calculations have been performed in a $12^{3}$ ${\bf k}$-point grid within the generalized gradient approximation of density functional theory \cite{PerdewPRL1996}, using norm-conserving pseudopotentials of the Goedecker-Hartwigsen-Hutter-Teter table \cite{HartwigsenPRB1998,*GoedeckerPRB1996}.
Phonon properties and electron-phonon matrix elements have been computed on a coarse $4^{3}$ ${\bf q}$-point grid,
and later interpolated to the triangular vertices forming the FS by the Wannier interpolation method.
A tessellation consisting of $\sim 7.5 \times 10^{3}$ vertices has been needed in this case to obtain converged results.

Figure \ref{fig:Fig5}(a) shows our results for the anisotropic mass-enhancement parameter on the six FS sheets of YH$_6$ at 300Gpa.
As can be appreciated in this figure, $\lambda_{{\bf k}}$ varies considerably among the different FS sheets, ranging from $\sim 1.0$ on the small electron pockets to $\sim 1.9$ on some regions of the biggest sheet.
Most importantly, $\lambda_{{\bf k}}$ varies substantially within the second and third sheets,
which in turn show strongly anisotropic and intricate topologies,
serving as a challenging test for our method.
We obtain a FS averaged mass-enhancement parameter of $\lambda=1.5$,
somewhat smaller than the $\lambda=1.9$ reported in Ref.~\cite{HeilPRB2019}.
We ascribe the discrepancy to the different FS integration method,
and note that our high-quality triangulated mesh gives a superconducting transition temperature which is in better agreement with the experimental results \cite{TroyanARXIV2019,KongARXIV2019}, as discussed in Ref.~\cite{AccompanyingPaper}.

We present in Figs.~\ref{fig:Fig5}(b) and (c) the first four fully symmetric HFSH functions of the second and third FS sheets, respectively.
Their corresponding $\lambda_{\tilde{L}}$ parameters are shown in Figs.~\ref{fig:Fig5}(d) and (e), respectively.
Similar to HEX-MgB$_2$, here we observe that the values of the coefficients decay very rapidly in this case as well.
This means that a relative accuracy of $\sim10^{-2}$ can be obtained in the description of the anisotropy of $\lambda_{{\bf k}}$ with less than ten coefficients in both sheets.
This results demonstrates that our methodology is equally valid for systems with any kind of symmetry or FS topology, and that it appears remarkably beneficial even in extremely anisotropic scenarios.


\section{Conclusions} \label{sec:concl}

In summary, we have presented a method to describe anisotropic Fermi surface quantities very efficiently.
This work constitutes an improvement over the HFSH basis set presented in Ref.~\cite{EigurenNJP2014}.
The major advance is the incorporation of crystal symmetries through the construction of a fully symmetric triangulated Fermi surface.
We have shown the general applicability of the method in systems with different symmetries.

As an application, we have demonstrated that the method is extremely efficient for compressing quantities related to the electron-phonon interaction in prototypical anisotropic superconductors.
The full potential of the method will be further illustrated in an accompanying paper \cite{AccompanyingPaper}, in which it is shown that the fully anisotropic Eliashberg equations of superconductivity can be solved in a very efficient and physically meaningful way in the HFSH representation.
In the case of conventional $s$-wave superconductors, only the fully symmetric HFSH subset introduced in this paper is needed.

Besides computational time and memory saving advantages, we believe that this work opens a path towards a quantitative comparison between different calculations ---and ultimately with experiments--- able to capture anisotropic effects.
Further ahead, we anticipate that this method can provide a tabulation of coefficients describing anisotropic physical quantities that could be included in material databases.
The need for anisotropy descriptors in the prediction of the superconducting critical temperature through machine-learning algorithms has been recently pointed out, for instance, in Ref.~\cite{XiePRB2019}.
We believe that the coefficients obtained through the procedure presented in this work are perfect candidates to be used as such descriptors.



\begin{acknowledgments}
The authors acknowledge the Department of Education, Universities and Research of the Basque Government
and the University of the Basque Country UPV/EHU (Grant No. IT756-13),
the Spanish Ministry of Economy and Competitiveness MINECO (Grants No. FIS2016-75862-P and No. PID2019-103910GB-I00)
and the University of the Basque Country UPV/EHU (Grant No. GIU18/138) for financial support.
J.L.-B. acknowledges the University of the Basque Country UPV/EHU (Grant No. PIF/UPV/16/240)
and the Donostia International Physics Center (DIPC) for financial support.
Computer facilities were provided by the DIPC.
\end{acknowledgments}

\bibliography{bibliography}

\begin{thebibliography}{56}%
\makeatletter
\providecommand \@ifxundefined [1]{%
 \@ifx{#1\undefined}
}%
\providecommand \@ifnum [1]{%
 \ifnum #1\expandafter \@firstoftwo
 \else \expandafter \@secondoftwo
 \fi
}%
\providecommand \@ifx [1]{%
 \ifx #1\expandafter \@firstoftwo
 \else \expandafter \@secondoftwo
 \fi
}%
\providecommand \natexlab [1]{#1}%
\providecommand \enquote  [1]{``#1''}%
\providecommand \bibnamefont  [1]{#1}%
\providecommand \bibfnamefont [1]{#1}%
\providecommand \citenamefont [1]{#1}%
\providecommand \href@noop [0]{\@secondoftwo}%
\providecommand \href [0]{\begingroup \@sanitize@url \@href}%
\providecommand \@href[1]{\@@startlink{#1}\@@href}%
\providecommand \@@href[1]{\endgroup#1\@@endlink}%
\providecommand \@sanitize@url [0]{\catcode `\\12\catcode `\$12\catcode
  `\&12\catcode `\#12\catcode `\^12\catcode `\_12\catcode `\%12\relax}%
\providecommand \@@startlink[1]{}%
\providecommand \@@endlink[0]{}%
\providecommand \url  [0]{\begingroup\@sanitize@url \@url }%
\providecommand \@url [1]{\endgroup\@href {#1}{\urlprefix }}%
\providecommand \urlprefix  [0]{URL }%
\providecommand \Eprint [0]{\href }%
\providecommand \doibase [0]{https://doi.org/}%
\providecommand \selectlanguage [0]{\@gobble}%
\providecommand \bibinfo  [0]{\@secondoftwo}%
\providecommand \bibfield  [0]{\@secondoftwo}%
\providecommand \translation [1]{[#1]}%
\providecommand \BibitemOpen [0]{}%
\providecommand \bibitemStop [0]{}%
\providecommand \bibitemNoStop [0]{.\EOS\space}%
\providecommand \EOS [0]{\spacefactor3000\relax}%
\providecommand \BibitemShut  [1]{\csname bibitem#1\endcsname}%
\let\auto@bib@innerbib\@empty
\bibitem [{\citenamefont {Grimvall}(1981)}]{Grimvall1981}%
  \BibitemOpen
  \bibfield  {author} {\bibinfo {author} {\bibfnamefont {G.}~\bibnamefont
  {Grimvall}},\ }\href@noop {} {\emph {\bibinfo {title} {The
  {E}lectron-{P}honon {I}nteraction in {M}etals}}}\ (\bibinfo  {publisher}
  {North-Holland, Amsterdam},\ \bibinfo {year} {1981})\BibitemShut {NoStop}%
\bibitem [{\citenamefont {Ziman}(1960)}]{Ziman1960}%
  \BibitemOpen
  \bibfield  {author} {\bibinfo {author} {\bibfnamefont {J.~M.}\ \bibnamefont
  {Ziman}},\ }\href@noop {} {\emph {\bibinfo {title} {Electrons and Phonons:
  The Theory of Transport Phenomena in Solids}}}\ (\bibinfo  {publisher}
  {Oxford University Press, Oxford},\ \bibinfo {year} {1960})\BibitemShut
  {NoStop}%
\bibitem [{\citenamefont {Schrieffer}(1964)}]{Schrieffer1964}%
  \BibitemOpen
  \bibfield  {author} {\bibinfo {author} {\bibfnamefont {J.~R.}\ \bibnamefont
  {Schrieffer}},\ }\href@noop {} {\emph {\bibinfo {title} {Theory of
  superconductivity}}}\ (\bibinfo  {publisher} {W.A. Benjamin, New York},\
  \bibinfo {year} {1964})\BibitemShut {NoStop}%
\bibitem [{\citenamefont {Allen}\ and\ \citenamefont
  {Mitrovic}(1983)}]{AllenMitrovic1983}%
  \BibitemOpen
  \bibfield  {author} {\bibinfo {author} {\bibfnamefont {P.~B.}\ \bibnamefont
  {Allen}}\ and\ \bibinfo {author} {\bibfnamefont {B.}~\bibnamefont
  {Mitrovic}},\ }\href
  {https://doi.org/https://doi.org/10.1016/S0081-1947(08)60665-7} {\emph
  {\bibinfo {title} {Theory of Superconducting {$T_{\rm c}$}}}},\ edited by\
  \bibinfo {editor} {\bibfnamefont {F.}~\bibnamefont {Seitz}}, \bibinfo
  {editor} {\bibfnamefont {D.}~\bibnamefont {Turnbull}},\ and\ \bibinfo
  {editor} {\bibfnamefont {H.}~\bibnamefont {Ehrenreich}},\ \bibinfo {series}
  {Solid State Physics}, Vol.~\bibinfo {volume} {37}\ (\bibinfo  {publisher}
  {Academic, New York},\ \bibinfo {year} {1983})\ pp.\ \bibinfo {pages}
  {1--92}\BibitemShut {NoStop}%
\bibitem [{\citenamefont {Giustino}(2017)}]{GiustinoRMP2016}%
  \BibitemOpen
  \bibfield  {author} {\bibinfo {author} {\bibfnamefont {F.}~\bibnamefont
  {Giustino}},\ }\bibfield  {title} {\bibinfo {title} {Electron-phonon
  interactions from first principles},\ }\href
  {https://doi.org/10.1103/RevModPhys.89.015003} {\bibfield  {journal}
  {\bibinfo  {journal} {Rev. Mod. Phys.}\ }\textbf {\bibinfo {volume} {89}},\
  \bibinfo {pages} {015003} (\bibinfo {year} {2017})}\BibitemShut {NoStop}%
\bibitem [{\citenamefont {Baroni}\ \emph {et~al.}(2001)\citenamefont {Baroni},
  \citenamefont {de~Gironcoli}, \citenamefont {Dal~Corso},\ and\ \citenamefont
  {Giannozzi}}]{BaroniRMP2001}%
  \BibitemOpen
  \bibfield  {author} {\bibinfo {author} {\bibfnamefont {S.}~\bibnamefont
  {Baroni}}, \bibinfo {author} {\bibfnamefont {S.}~\bibnamefont
  {de~Gironcoli}}, \bibinfo {author} {\bibfnamefont {A.}~\bibnamefont
  {Dal~Corso}},\ and\ \bibinfo {author} {\bibfnamefont {P.}~\bibnamefont
  {Giannozzi}},\ }\bibfield  {title} {\bibinfo {title} {Phonons and related
  crystal properties from density-functional perturbation theory},\ }\href
  {https://doi.org/10.1103/RevModPhys.73.515} {\bibfield  {journal} {\bibinfo
  {journal} {Rev. Mod. Phys.}\ }\textbf {\bibinfo {volume} {73}},\ \bibinfo
  {pages} {515} (\bibinfo {year} {2001})}\BibitemShut {NoStop}%
\bibitem [{\citenamefont {Giustino}\ \emph {et~al.}(2007)\citenamefont
  {Giustino}, \citenamefont {Cohen},\ and\ \citenamefont
  {Louie}}]{GiustinoPRB2007}%
  \BibitemOpen
  \bibfield  {author} {\bibinfo {author} {\bibfnamefont {F.}~\bibnamefont
  {Giustino}}, \bibinfo {author} {\bibfnamefont {M.~L.}\ \bibnamefont
  {Cohen}},\ and\ \bibinfo {author} {\bibfnamefont {S.~G.}\ \bibnamefont
  {Louie}},\ }\bibfield  {title} {\bibinfo {title} {Electron-phonon interaction
  using {W}annier functions},\ }\href
  {https://doi.org/10.1103/PhysRevB.76.165108} {\bibfield  {journal} {\bibinfo
  {journal} {Phys. Rev. B}\ }\textbf {\bibinfo {volume} {76}},\ \bibinfo
  {pages} {165108} (\bibinfo {year} {2007})}\BibitemShut {NoStop}%
\bibitem [{\citenamefont {Eiguren}\ and\ \citenamefont
  {Ambrosch-Draxl}(2008{\natexlab{a}})}]{EigurenPRB2008}%
  \BibitemOpen
  \bibfield  {author} {\bibinfo {author} {\bibfnamefont {A.}~\bibnamefont
  {Eiguren}}\ and\ \bibinfo {author} {\bibfnamefont {C.}~\bibnamefont
  {Ambrosch-Draxl}},\ }\bibfield  {title} {\bibinfo {title} {Wannier
  interpolation scheme for phonon-induced potentials: Application to bulk
  {M}g{B}$_{2}$, {W}, and the $(1\ifmmode\times\else\texttimes\fi{}1)$
  {H}-covered {W}(110) surface},\ }\href
  {https://doi.org/10.1103/PhysRevB.78.045124} {\bibfield  {journal} {\bibinfo
  {journal} {Phys. Rev. B}\ }\textbf {\bibinfo {volume} {78}},\ \bibinfo
  {pages} {045124} (\bibinfo {year} {2008}{\natexlab{a}})}\BibitemShut
  {NoStop}%
\bibitem [{\citenamefont {Mustafa}\ \emph {et~al.}(2016)\citenamefont
  {Mustafa}, \citenamefont {Bernardi}, \citenamefont {Neaton},\ and\
  \citenamefont {Louie}}]{MustafaPRB2016}%
  \BibitemOpen
  \bibfield  {author} {\bibinfo {author} {\bibfnamefont {J.~I.}\ \bibnamefont
  {Mustafa}}, \bibinfo {author} {\bibfnamefont {M.}~\bibnamefont {Bernardi}},
  \bibinfo {author} {\bibfnamefont {J.~B.}\ \bibnamefont {Neaton}},\ and\
  \bibinfo {author} {\bibfnamefont {S.~G.}\ \bibnamefont {Louie}},\ }\bibfield
  {title} {\bibinfo {title} {Ab initio electronic relaxation times and
  transport in noble metals},\ }\href
  {https://doi.org/10.1103/PhysRevB.94.155105} {\bibfield  {journal} {\bibinfo
  {journal} {Phys. Rev. B}\ }\textbf {\bibinfo {volume} {94}},\ \bibinfo
  {pages} {155105} (\bibinfo {year} {2016})}\BibitemShut {NoStop}%
\bibitem [{\citenamefont {Ponc\'e}\ \emph {et~al.}(2018)\citenamefont
  {Ponc\'e}, \citenamefont {Margine},\ and\ \citenamefont
  {Giustino}}]{PoncePRB2018}%
  \BibitemOpen
  \bibfield  {author} {\bibinfo {author} {\bibfnamefont {S.}~\bibnamefont
  {Ponc\'e}}, \bibinfo {author} {\bibfnamefont {E.~R.}\ \bibnamefont
  {Margine}},\ and\ \bibinfo {author} {\bibfnamefont {F.}~\bibnamefont
  {Giustino}},\ }\bibfield  {title} {\bibinfo {title} {Towards predictive
  many-body calculations of phonon-limited carrier mobilities in
  semiconductors},\ }\href {https://doi.org/10.1103/PhysRevB.97.121201}
  {\bibfield  {journal} {\bibinfo  {journal} {Phys. Rev. B}\ }\textbf {\bibinfo
  {volume} {97}},\ \bibinfo {pages} {121201(R)} (\bibinfo {year}
  {2018})}\BibitemShut {NoStop}%
\bibitem [{\citenamefont {Calandra}\ \emph {et~al.}(2010)\citenamefont
  {Calandra}, \citenamefont {Profeta},\ and\ \citenamefont
  {Mauri}}]{CalandraPRB2010}%
  \BibitemOpen
  \bibfield  {author} {\bibinfo {author} {\bibfnamefont {M.}~\bibnamefont
  {Calandra}}, \bibinfo {author} {\bibfnamefont {G.}~\bibnamefont {Profeta}},\
  and\ \bibinfo {author} {\bibfnamefont {F.}~\bibnamefont {Mauri}},\ }\bibfield
   {title} {\bibinfo {title} {Adiabatic and nonadiabatic phonon dispersion in a
  {W}annier function approach},\ }\href
  {https://doi.org/10.1103/PhysRevB.82.165111} {\bibfield  {journal} {\bibinfo
  {journal} {Phys. Rev. B}\ }\textbf {\bibinfo {volume} {82}},\ \bibinfo
  {pages} {165111} (\bibinfo {year} {2010})}\BibitemShut {NoStop}%
\bibitem [{\citenamefont {Novko}(2018)}]{NovkoPRB2018}%
  \BibitemOpen
  \bibfield  {author} {\bibinfo {author} {\bibfnamefont {D.}~\bibnamefont
  {Novko}},\ }\bibfield  {title} {\bibinfo {title} {Nonadiabatic coupling
  effects in {M}g{B}$_{2}$ reexamined},\ }\href
  {https://doi.org/10.1103/PhysRevB.98.041112} {\bibfield  {journal} {\bibinfo
  {journal} {Phys. Rev. B}\ }\textbf {\bibinfo {volume} {98}},\ \bibinfo
  {pages} {041112(R)} (\bibinfo {year} {2018})}\BibitemShut {NoStop}%
\bibitem [{\citenamefont {Garcia-Goiricelaya}\ \emph
  {et~al.}(2020)\citenamefont {Garcia-Goiricelaya}, \citenamefont
  {Lafuente-Bartolome}, \citenamefont {Gurtubay},\ and\ \citenamefont
  {Eiguren}}]{GoiricelayaPRB2020}%
  \BibitemOpen
  \bibfield  {author} {\bibinfo {author} {\bibfnamefont {P.}~\bibnamefont
  {Garcia-Goiricelaya}}, \bibinfo {author} {\bibfnamefont {J.}~\bibnamefont
  {Lafuente-Bartolome}}, \bibinfo {author} {\bibfnamefont {I.~G.}\ \bibnamefont
  {Gurtubay}},\ and\ \bibinfo {author} {\bibfnamefont {A.}~\bibnamefont
  {Eiguren}},\ }\bibfield  {title} {\bibinfo {title} {Emergence of large
  nonadiabatic effects induced by the electron-phonon interaction on the
  complex vibrational quasiparticle spectrum of doped monolayer
  {M}o{S}$_{2}$},\ }\href {https://doi.org/10.1103/PhysRevB.101.054304}
  {\bibfield  {journal} {\bibinfo  {journal} {Phys. Rev. B}\ }\textbf {\bibinfo
  {volume} {101}},\ \bibinfo {pages} {054304} (\bibinfo {year}
  {2020})}\BibitemShut {NoStop}%
\bibitem [{\citenamefont {Eiguren}\ \emph {et~al.}(2003)\citenamefont
  {Eiguren}, \citenamefont {de~Gironcoli}, \citenamefont {Chulkov},
  \citenamefont {Echenique},\ and\ \citenamefont {Tosatti}}]{EigurenPRL2003}%
  \BibitemOpen
  \bibfield  {author} {\bibinfo {author} {\bibfnamefont {A.}~\bibnamefont
  {Eiguren}}, \bibinfo {author} {\bibfnamefont {S.}~\bibnamefont
  {de~Gironcoli}}, \bibinfo {author} {\bibfnamefont {E.~V.}\ \bibnamefont
  {Chulkov}}, \bibinfo {author} {\bibfnamefont {P.~M.}\ \bibnamefont
  {Echenique}},\ and\ \bibinfo {author} {\bibfnamefont {E.}~\bibnamefont
  {Tosatti}},\ }\bibfield  {title} {\bibinfo {title} {Electron-phonon
  interaction at the {B}e(0001) surface},\ }\href
  {https://doi.org/10.1103/PhysRevLett.91.166803} {\bibfield  {journal}
  {\bibinfo  {journal} {Phys. Rev. Lett.}\ }\textbf {\bibinfo {volume} {91}},\
  \bibinfo {pages} {166803} (\bibinfo {year} {2003})}\BibitemShut {NoStop}%
\bibitem [{\citenamefont {Verdi}\ \emph {et~al.}(2017)\citenamefont {Verdi},
  \citenamefont {Caruso},\ and\ \citenamefont {Giustino}}]{VerdiNCM2017}%
  \BibitemOpen
  \bibfield  {author} {\bibinfo {author} {\bibfnamefont {C.}~\bibnamefont
  {Verdi}}, \bibinfo {author} {\bibfnamefont {F.}~\bibnamefont {Caruso}},\ and\
  \bibinfo {author} {\bibfnamefont {F.}~\bibnamefont {Giustino}},\ }\bibfield
  {title} {\bibinfo {title} {Origin of the crossover from polarons to fermi
  liquids in transition metal oxides},\ }\href
  {https://doi.org/10.1038/ncomms15769} {\bibfield  {journal} {\bibinfo
  {journal} {Nature Communications}\ }\textbf {\bibinfo {volume} {8}},\
  \bibinfo {pages} {15769} (\bibinfo {year} {2017})}\BibitemShut {NoStop}%
\bibitem [{\citenamefont {Garcia-Goiricelaya}\ \emph
  {et~al.}(2019)\citenamefont {Garcia-Goiricelaya}, \citenamefont
  {Lafuente-Bartolome}, \citenamefont {Gurtubay},\ and\ \citenamefont
  {Eiguren}}]{GoiricelayaCMP2019}%
  \BibitemOpen
  \bibfield  {author} {\bibinfo {author} {\bibfnamefont {P.}~\bibnamefont
  {Garcia-Goiricelaya}}, \bibinfo {author} {\bibfnamefont {J.}~\bibnamefont
  {Lafuente-Bartolome}}, \bibinfo {author} {\bibfnamefont {I.~G.}\ \bibnamefont
  {Gurtubay}},\ and\ \bibinfo {author} {\bibfnamefont {A.}~\bibnamefont
  {Eiguren}},\ }\bibfield  {title} {\bibinfo {title} {Long-living carriers in a
  strong electron-phonon interacting two-dimensional doped semiconductor},\
  }\href {https://doi.org/10.1038/s42005-019-0182-0} {\bibfield  {journal}
  {\bibinfo  {journal} {Communications Physics}\ }\textbf {\bibinfo {volume}
  {2}},\ \bibinfo {pages} {81} (\bibinfo {year} {2019})}\BibitemShut {NoStop}%
\bibitem [{\citenamefont {Margine}\ and\ \citenamefont
  {Giustino}(2013)}]{MarginePRB2013}%
  \BibitemOpen
  \bibfield  {author} {\bibinfo {author} {\bibfnamefont {E.~R.}\ \bibnamefont
  {Margine}}\ and\ \bibinfo {author} {\bibfnamefont {F.}~\bibnamefont
  {Giustino}},\ }\bibfield  {title} {\bibinfo {title} {Anisotropic
  {M}igdal-{E}liashberg theory using {W}annier functions},\ }\href
  {https://doi.org/10.1103/PhysRevB.87.024505} {\bibfield  {journal} {\bibinfo
  {journal} {Phys. Rev. B}\ }\textbf {\bibinfo {volume} {87}},\ \bibinfo
  {pages} {024505} (\bibinfo {year} {2013})}\BibitemShut {NoStop}%
\bibitem [{\citenamefont {Allen}(1976)}]{AllenPRB1975}%
  \BibitemOpen
  \bibfield  {author} {\bibinfo {author} {\bibfnamefont {P.~B.}\ \bibnamefont
  {Allen}},\ }\bibfield  {title} {\bibinfo {title} {Fermi-surface harmonics: A
  general method for nonspherical problems. application to {B}oltzmann and
  {E}liashberg equations},\ }\href {https://doi.org/10.1103/PhysRevB.13.1416}
  {\bibfield  {journal} {\bibinfo  {journal} {Phys. Rev. B}\ }\textbf {\bibinfo
  {volume} {13}},\ \bibinfo {pages} {1416} (\bibinfo {year}
  {1976})}\BibitemShut {NoStop}%
\bibitem [{\citenamefont {Butler}\ and\ \citenamefont
  {Allen}(1976)}]{Butler1976}%
  \BibitemOpen
  \bibfield  {author} {\bibinfo {author} {\bibfnamefont {W.~H.}\ \bibnamefont
  {Butler}}\ and\ \bibinfo {author} {\bibfnamefont {P.~B.}\ \bibnamefont
  {Allen}},\ }\bibinfo {title} {Gap anisotropy and {T}c enhancement: General
  theory, and calculations for {N}b, using {F}ermi surface harmonics},\ in\
  \href {https://doi.org/10.1007/978-1-4615-8795-8_6} {\emph {\bibinfo
  {booktitle} {Superconductivity in d- and f-Band Metals: Second Rochester
  Conference}}},\ \bibinfo {editor} {edited by\ \bibinfo {editor}
  {\bibfnamefont {D.~H.}\ \bibnamefont {Douglass}}}\ (\bibinfo  {publisher}
  {Springer US},\ \bibinfo {address} {Boston, MA},\ \bibinfo {year} {1976})\
  pp.\ \bibinfo {pages} {73--120}\BibitemShut {NoStop}%
\bibitem [{\citenamefont {Heid}\ \emph {et~al.}(2008)\citenamefont {Heid},
  \citenamefont {Bohnen}, \citenamefont {Zeyher},\ and\ \citenamefont
  {Manske}}]{HeidPRL2008}%
  \BibitemOpen
  \bibfield  {author} {\bibinfo {author} {\bibfnamefont {R.}~\bibnamefont
  {Heid}}, \bibinfo {author} {\bibfnamefont {K.-P.}\ \bibnamefont {Bohnen}},
  \bibinfo {author} {\bibfnamefont {R.}~\bibnamefont {Zeyher}},\ and\ \bibinfo
  {author} {\bibfnamefont {D.}~\bibnamefont {Manske}},\ }\bibfield  {title}
  {\bibinfo {title} {Momentum dependence of the electron-phonon coupling and
  self-energy effects in superconducting {Y}{B}a$_{2}${C}u$_{3}${O}$_{7}$
  within the local density approximation},\ }\href
  {https://doi.org/10.1103/PhysRevLett.100.137001} {\bibfield  {journal}
  {\bibinfo  {journal} {Phys. Rev. Lett.}\ }\textbf {\bibinfo {volume} {100}},\
  \bibinfo {pages} {137001} (\bibinfo {year} {2008})}\BibitemShut {NoStop}%
\bibitem [{\citenamefont {Xu}\ and\ \citenamefont
  {Verstraete}(2014)}]{XuPRL2014}%
  \BibitemOpen
  \bibfield  {author} {\bibinfo {author} {\bibfnamefont {B.}~\bibnamefont
  {Xu}}\ and\ \bibinfo {author} {\bibfnamefont {M.~J.}\ \bibnamefont
  {Verstraete}},\ }\bibfield  {title} {\bibinfo {title} {First principles
  explanation of the positive seebeck coefficient of lithium},\ }\href
  {https://doi.org/10.1103/PhysRevLett.112.196603} {\bibfield  {journal}
  {\bibinfo  {journal} {Phys. Rev. Lett.}\ }\textbf {\bibinfo {volume} {112}},\
  \bibinfo {pages} {196603} (\bibinfo {year} {2014})}\BibitemShut {NoStop}%
\bibitem [{\citenamefont {Eiguren}\ and\ \citenamefont
  {Gurtubay}(2014)}]{EigurenNJP2014}%
  \BibitemOpen
  \bibfield  {author} {\bibinfo {author} {\bibfnamefont {A.}~\bibnamefont
  {Eiguren}}\ and\ \bibinfo {author} {\bibfnamefont {I.~G.}\ \bibnamefont
  {Gurtubay}},\ }\bibfield  {title} {\bibinfo {title} {Helmholtz {F}ermi
  surface harmonics: an efficient approach for treating anisotropic problems
  involving {F}ermi surface integrals},\ }\href
  {https://doi.org/10.1088/1367-2630/16/6/063014} {\bibfield  {journal}
  {\bibinfo  {journal} {New Journal of Physics}\ }\textbf {\bibinfo {volume}
  {16}},\ \bibinfo {pages} {063014} (\bibinfo {year} {2014})}\BibitemShut
  {NoStop}%
\bibitem [{\citenamefont {Bl\"ochl}\ \emph {et~al.}(1994)\citenamefont
  {Bl\"ochl}, \citenamefont {Jepsen},\ and\ \citenamefont
  {Andersen}}]{BlochlPRB1994}%
  \BibitemOpen
  \bibfield  {author} {\bibinfo {author} {\bibfnamefont {P.~E.}\ \bibnamefont
  {Bl\"ochl}}, \bibinfo {author} {\bibfnamefont {O.}~\bibnamefont {Jepsen}},\
  and\ \bibinfo {author} {\bibfnamefont {O.~K.}\ \bibnamefont {Andersen}},\
  }\bibfield  {title} {\bibinfo {title} {Improved tetrahedron method for
  {B}rillouin-zone integrations},\ }\href
  {https://doi.org/10.1103/PhysRevB.49.16223} {\bibfield  {journal} {\bibinfo
  {journal} {Phys. Rev. B}\ }\textbf {\bibinfo {volume} {49}},\ \bibinfo
  {pages} {16223} (\bibinfo {year} {1994})}\BibitemShut {NoStop}%
\bibitem [{\citenamefont {Aroyo}\ \emph {et~al.}(2014)\citenamefont {Aroyo},
  \citenamefont {Orobengoa}, \citenamefont {de~la Flor}, \citenamefont {Tasci},
  \citenamefont {Perez-Mato},\ and\ \citenamefont
  {Wondratschek}}]{AroyoAC2014}%
  \BibitemOpen
  \bibfield  {author} {\bibinfo {author} {\bibfnamefont {M.~I.}\ \bibnamefont
  {Aroyo}}, \bibinfo {author} {\bibfnamefont {D.}~\bibnamefont {Orobengoa}},
  \bibinfo {author} {\bibfnamefont {G.}~\bibnamefont {de~la Flor}}, \bibinfo
  {author} {\bibfnamefont {E.~S.}\ \bibnamefont {Tasci}}, \bibinfo {author}
  {\bibfnamefont {J.~M.}\ \bibnamefont {Perez-Mato}},\ and\ \bibinfo {author}
  {\bibfnamefont {H.}~\bibnamefont {Wondratschek}},\ }\bibfield  {title}
  {\bibinfo {title} {Brillouin-zone database on the {B}ilbao {C}rystallographic
  {S}erver},\ }\href {https://doi.org/10.1107/S205327331303091X} {\bibfield
  {journal} {\bibinfo  {journal} {Acta Crystallographica Section A}\ }\textbf
  {\bibinfo {volume} {70}},\ \bibinfo {pages} {126} (\bibinfo {year}
  {2014})}\BibitemShut {NoStop}%
\bibitem [{\citenamefont {Shewchuk}(1996)}]{ShewchukSpringer1996}%
  \BibitemOpen
  \bibfield  {author} {\bibinfo {author} {\bibfnamefont {J.~R.}\ \bibnamefont
  {Shewchuk}},\ }\bibfield  {title} {\bibinfo {title} {Triangle: {E}ngineering
  a {2D} {Q}uality {M}esh {G}enerator and {D}elaunay {T}riangulator},\ }in\
  \href@noop {} {\emph {\bibinfo {booktitle} {Applied Computational Geometry:
  Towards Geometric Engineering}}},\ \bibinfo {series} {Lecture Notes in
  Computer Science}, Vol.\ \bibinfo {volume} {1148},\ \bibinfo {editor} {edited
  by\ \bibinfo {editor} {\bibfnamefont {M.~C.}\ \bibnamefont {Lin}}\ and\
  \bibinfo {editor} {\bibfnamefont {D.}~\bibnamefont {Manocha}}}\ (\bibinfo
  {publisher} {Springer-Verlag, Berlin},\ \bibinfo {year} {1996})\ pp.\
  \bibinfo {pages} {203--222}\BibitemShut {NoStop}%
\bibitem [{\citenamefont {Si}(2015)}]{SiACM2015}%
  \BibitemOpen
  \bibfield  {author} {\bibinfo {author} {\bibfnamefont {H.}~\bibnamefont
  {Si}},\ }\bibfield  {title} {\bibinfo {title} {Tetgen, a delaunay-based
  quality tetrahedral mesh generator},\ }\bibfield  {journal} {\bibinfo
  {journal} {ACM Trans. Math. Softw.}\ }\textbf {\bibinfo {volume} {41}},\
  \href {https://doi.org/10.1145/2629697} {10.1145/2629697} (\bibinfo {year}
  {2015})\BibitemShut {NoStop}%
\bibitem [{Note1()}]{Note1}%
  \BibitemOpen
  \bibinfo {note} {The ground state calculations for FCC-Cu are performed
  within the generalized gradient approximation of density functional theory
  \cite {PerdewPRL1996} on a coarse $12\times 12\times 12$ ${\protect \bf
  k}$-point grid using the {\protect \sc Quantum ESPRESSO} package \cite
  {QE2017}, and the energies at the tetrahedral vertices are obtained by means
  of Wannier interpolation \cite
  {MarzariPRB1997,SouzaPRB2001,PizziJPCM2020}.}\BibitemShut {Stop}%
\bibitem [{\citenamefont {Botsch}\ \emph {et~al.}(2010)\citenamefont {Botsch},
  \citenamefont {Kobbelt}, \citenamefont {Pauly}, \citenamefont {Alliez},\ and\
  \citenamefont {L{\'e}vy}}]{botschCRC2010}%
  \BibitemOpen
  \bibfield  {author} {\bibinfo {author} {\bibfnamefont {M.}~\bibnamefont
  {Botsch}}, \bibinfo {author} {\bibfnamefont {L.}~\bibnamefont {Kobbelt}},
  \bibinfo {author} {\bibfnamefont {M.}~\bibnamefont {Pauly}}, \bibinfo
  {author} {\bibfnamefont {P.}~\bibnamefont {Alliez}},\ and\ \bibinfo {author}
  {\bibfnamefont {B.}~\bibnamefont {L{\'e}vy}},\ }\href@noop {} {\emph
  {\bibinfo {title} {Polygon mesh processing}}}\ (\bibinfo  {publisher} {AK
  Peters/CRC Press, New York},\ \bibinfo {year} {2010})\BibitemShut {NoStop}%
\bibitem [{\citenamefont {Marzari}\ and\ \citenamefont
  {Vanderbilt}(1997)}]{MarzariPRB1997}%
  \BibitemOpen
  \bibfield  {author} {\bibinfo {author} {\bibfnamefont {N.}~\bibnamefont
  {Marzari}}\ and\ \bibinfo {author} {\bibfnamefont {D.}~\bibnamefont
  {Vanderbilt}},\ }\bibfield  {title} {\bibinfo {title} {Maximally localized
  generalized {W}annier functions for composite energy bands},\ }\href
  {https://doi.org/10.1103/PhysRevB.56.12847} {\bibfield  {journal} {\bibinfo
  {journal} {Phys. Rev. B}\ }\textbf {\bibinfo {volume} {56}},\ \bibinfo
  {pages} {12847} (\bibinfo {year} {1997})}\BibitemShut {NoStop}%
\bibitem [{\citenamefont {Souza}\ \emph {et~al.}(2001)\citenamefont {Souza},
  \citenamefont {Marzari},\ and\ \citenamefont {Vanderbilt}}]{SouzaPRB2001}%
  \BibitemOpen
  \bibfield  {author} {\bibinfo {author} {\bibfnamefont {I.}~\bibnamefont
  {Souza}}, \bibinfo {author} {\bibfnamefont {N.}~\bibnamefont {Marzari}},\
  and\ \bibinfo {author} {\bibfnamefont {D.}~\bibnamefont {Vanderbilt}},\
  }\bibfield  {title} {\bibinfo {title} {Maximally localized {W}annier
  functions for entangled energy bands},\ }\href
  {https://doi.org/10.1103/PhysRevB.65.035109} {\bibfield  {journal} {\bibinfo
  {journal} {Phys. Rev. B}\ }\textbf {\bibinfo {volume} {65}},\ \bibinfo
  {pages} {035109} (\bibinfo {year} {2001})}\BibitemShut {NoStop}%
\bibitem [{\citenamefont {Pizzi}\ \emph {et~al.}(2020)\citenamefont {Pizzi},
  \citenamefont {Vitale}, \citenamefont {Arita}, \citenamefont {BlÃ¼gel},
  \citenamefont {Freimuth}, \citenamefont {G{\'{e}}ranton}, \citenamefont
  {Gibertini}, \citenamefont {Gresch}, \citenamefont {Johnson}, \citenamefont
  {Koretsune}, \citenamefont {Iba{\~{n}}ez-Azpiroz}, \citenamefont {Lee},
  \citenamefont {Lihm}, \citenamefont {Marchand}, \citenamefont {Marrazzo},
  \citenamefont {Mokrousov}, \citenamefont {Mustafa}, \citenamefont {Nohara},
  \citenamefont {Nomura}, \citenamefont {Paulatto}, \citenamefont
  {Ponc{\'{e}}}, \citenamefont {Ponweiser}, \citenamefont {Qiao}, \citenamefont
  {ThÃ¶le}, \citenamefont {Tsirkin}, \citenamefont {Wierzbowska},
  \citenamefont {Marzari}, \citenamefont {Vanderbilt}, \citenamefont {Souza},
  \citenamefont {Mostofi},\ and\ \citenamefont {Yates}}]{PizziJPCM2020}%
  \BibitemOpen
  \bibfield  {author} {\bibinfo {author} {\bibfnamefont {G.}~\bibnamefont
  {Pizzi}}, \bibinfo {author} {\bibfnamefont {V.}~\bibnamefont {Vitale}},
  \bibinfo {author} {\bibfnamefont {R.}~\bibnamefont {Arita}}, \bibinfo
  {author} {\bibfnamefont {S.}~\bibnamefont {BlÃ¼gel}}, \bibinfo {author}
  {\bibfnamefont {F.}~\bibnamefont {Freimuth}}, \bibinfo {author}
  {\bibfnamefont {G.}~\bibnamefont {G{\'{e}}ranton}}, \bibinfo {author}
  {\bibfnamefont {M.}~\bibnamefont {Gibertini}}, \bibinfo {author}
  {\bibfnamefont {D.}~\bibnamefont {Gresch}}, \bibinfo {author} {\bibfnamefont
  {C.}~\bibnamefont {Johnson}}, \bibinfo {author} {\bibfnamefont
  {T.}~\bibnamefont {Koretsune}}, \bibinfo {author} {\bibfnamefont
  {J.}~\bibnamefont {Iba{\~{n}}ez-Azpiroz}}, \bibinfo {author} {\bibfnamefont
  {H.}~\bibnamefont {Lee}}, \bibinfo {author} {\bibfnamefont {J.-M.}\
  \bibnamefont {Lihm}}, \bibinfo {author} {\bibfnamefont {D.}~\bibnamefont
  {Marchand}}, \bibinfo {author} {\bibfnamefont {A.}~\bibnamefont {Marrazzo}},
  \bibinfo {author} {\bibfnamefont {Y.}~\bibnamefont {Mokrousov}}, \bibinfo
  {author} {\bibfnamefont {J.~I.}\ \bibnamefont {Mustafa}}, \bibinfo {author}
  {\bibfnamefont {Y.}~\bibnamefont {Nohara}}, \bibinfo {author} {\bibfnamefont
  {Y.}~\bibnamefont {Nomura}}, \bibinfo {author} {\bibfnamefont
  {L.}~\bibnamefont {Paulatto}}, \bibinfo {author} {\bibfnamefont
  {S.}~\bibnamefont {Ponc{\'{e}}}}, \bibinfo {author} {\bibfnamefont
  {T.}~\bibnamefont {Ponweiser}}, \bibinfo {author} {\bibfnamefont
  {J.}~\bibnamefont {Qiao}}, \bibinfo {author} {\bibfnamefont {F.}~\bibnamefont
  {ThÃ¶le}}, \bibinfo {author} {\bibfnamefont {S.~S.}\ \bibnamefont
  {Tsirkin}}, \bibinfo {author} {\bibfnamefont {M.}~\bibnamefont
  {Wierzbowska}}, \bibinfo {author} {\bibfnamefont {N.}~\bibnamefont
  {Marzari}}, \bibinfo {author} {\bibfnamefont {D.}~\bibnamefont {Vanderbilt}},
  \bibinfo {author} {\bibfnamefont {I.}~\bibnamefont {Souza}}, \bibinfo
  {author} {\bibfnamefont {A.~A.}\ \bibnamefont {Mostofi}},\ and\ \bibinfo
  {author} {\bibfnamefont {J.~R.}\ \bibnamefont {Yates}},\ }\bibfield  {title}
  {\bibinfo {title} {Wannier90 as a community code: new features and
  applications},\ }\href {https://doi.org/10.1088/1361-648x/ab51ff} {\bibfield
  {journal} {\bibinfo  {journal} {Journal of Physics: Condensed Matter}\
  }\textbf {\bibinfo {volume} {32}},\ \bibinfo {pages} {165902} (\bibinfo
  {year} {2020})}\BibitemShut {NoStop}%
\bibitem [{\citenamefont {Kawamura}(2019)}]{KawamuraCPC2019}%
  \BibitemOpen
  \bibfield  {author} {\bibinfo {author} {\bibfnamefont {M.}~\bibnamefont
  {Kawamura}},\ }\bibfield  {title} {\bibinfo {title} {Fermi{S}urfer:
  {F}ermi-surface viewer providing multiple representation schemes},\ }\href
  {https://doi.org/https://doi.org/10.1016/j.cpc.2019.01.017} {\bibfield
  {journal} {\bibinfo  {journal} {Computer Physics Communications}\ }\textbf
  {\bibinfo {volume} {239}},\ \bibinfo {pages} {197 } (\bibinfo {year}
  {2019})}\BibitemShut {NoStop}%
\bibitem [{\citenamefont {Rittweger}\ \emph {et~al.}(2017)\citenamefont
  {Rittweger}, \citenamefont {Hinsche},\ and\ \citenamefont
  {Mertig}}]{RittwegerJPCM2017}%
  \BibitemOpen
  \bibfield  {author} {\bibinfo {author} {\bibfnamefont {F.}~\bibnamefont
  {Rittweger}}, \bibinfo {author} {\bibfnamefont {N.~F.}\ \bibnamefont
  {Hinsche}},\ and\ \bibinfo {author} {\bibfnamefont {I.}~\bibnamefont
  {Mertig}},\ }\bibfield  {title} {\bibinfo {title} {Phonon limited electronic
  transport in {P}b},\ }\href {https://doi.org/10.1088/1361-648x/aa7b56}
  {\bibfield  {journal} {\bibinfo  {journal} {Journal of Physics: Condensed
  Matter}\ }\textbf {\bibinfo {volume} {29}},\ \bibinfo {pages} {355501}
  (\bibinfo {year} {2017})}\BibitemShut {NoStop}%
\bibitem [{\citenamefont {McMillan}(1968)}]{McMillanPR1968}%
  \BibitemOpen
  \bibfield  {author} {\bibinfo {author} {\bibfnamefont {W.~L.}\ \bibnamefont
  {McMillan}},\ }\bibfield  {title} {\bibinfo {title} {Transition temperature
  of strong-coupled superconductors},\ }\href
  {https://doi.org/10.1103/PhysRev.167.331} {\bibfield  {journal} {\bibinfo
  {journal} {Phys. Rev.}\ }\textbf {\bibinfo {volume} {167}},\ \bibinfo {pages}
  {331} (\bibinfo {year} {1968})}\BibitemShut {NoStop}%
\bibitem [{\citenamefont {Allen}\ and\ \citenamefont
  {Dynes}(1975)}]{AllenDynesPRB1975}%
  \BibitemOpen
  \bibfield  {author} {\bibinfo {author} {\bibfnamefont {P.~B.}\ \bibnamefont
  {Allen}}\ and\ \bibinfo {author} {\bibfnamefont {R.~C.}\ \bibnamefont
  {Dynes}},\ }\bibfield  {title} {\bibinfo {title} {Transition temperature of
  strong-coupled superconductors reanalyzed},\ }\href
  {https://doi.org/10.1103/PhysRevB.12.905} {\bibfield  {journal} {\bibinfo
  {journal} {Phys. Rev. B}\ }\textbf {\bibinfo {volume} {12}},\ \bibinfo
  {pages} {905} (\bibinfo {year} {1975})}\BibitemShut {NoStop}%
\bibitem [{\citenamefont {Giannozzi}\ \emph {et~al.}(2017)\citenamefont
  {Giannozzi}, \citenamefont {Andreussi}, \citenamefont {Brumme}, \citenamefont
  {Bunau}, \citenamefont {Nardelli}, \citenamefont {Calandra}, \citenamefont
  {Car}, \citenamefont {Cavazzoni}, \citenamefont {Ceresoli}, \citenamefont
  {Cococcioni}, \citenamefont {Colonna}, \citenamefont {Carnimeo},
  \citenamefont {Corso}, \citenamefont {de~Gironcoli}, \citenamefont {Delugas},
  \citenamefont {Jr}, \citenamefont {Ferretti}, \citenamefont {Floris},
  \citenamefont {Fratesi}, \citenamefont {Fugallo}, \citenamefont {Gebauer},
  \citenamefont {Gerstmann}, \citenamefont {Giustino}, \citenamefont {Gorni},
  \citenamefont {Jia}, \citenamefont {Kawamura}, \citenamefont {Ko},
  \citenamefont {Kokalj}, \citenamefont {Küçükbenli}, \citenamefont
  {Lazzeri}, \citenamefont {Marsili}, \citenamefont {Marzari}, \citenamefont
  {Mauri}, \citenamefont {Nguyen}, \citenamefont {Nguyen}, \citenamefont {de-la
  Roza}, \citenamefont {Paulatto}, \citenamefont {Poncé}, \citenamefont
  {Rocca}, \citenamefont {Sabatini}, \citenamefont {Santra}, \citenamefont
  {Schlipf}, \citenamefont {Seitsonen}, \citenamefont {Smogunov}, \citenamefont
  {Timrov}, \citenamefont {Thonhauser}, \citenamefont {Umari}, \citenamefont
  {Vast}, \citenamefont {Wu},\ and\ \citenamefont {Baroni}}]{QE2017}%
  \BibitemOpen
  \bibfield  {author} {\bibinfo {author} {\bibfnamefont {P.}~\bibnamefont
  {Giannozzi}}, \bibinfo {author} {\bibfnamefont {O.}~\bibnamefont
  {Andreussi}}, \bibinfo {author} {\bibfnamefont {T.}~\bibnamefont {Brumme}},
  \bibinfo {author} {\bibfnamefont {O.}~\bibnamefont {Bunau}}, \bibinfo
  {author} {\bibfnamefont {M.~B.}\ \bibnamefont {Nardelli}}, \bibinfo {author}
  {\bibfnamefont {M.}~\bibnamefont {Calandra}}, \bibinfo {author}
  {\bibfnamefont {R.}~\bibnamefont {Car}}, \bibinfo {author} {\bibfnamefont
  {C.}~\bibnamefont {Cavazzoni}}, \bibinfo {author} {\bibfnamefont
  {D.}~\bibnamefont {Ceresoli}}, \bibinfo {author} {\bibfnamefont
  {M.}~\bibnamefont {Cococcioni}}, \bibinfo {author} {\bibfnamefont
  {N.}~\bibnamefont {Colonna}}, \bibinfo {author} {\bibfnamefont
  {I.}~\bibnamefont {Carnimeo}}, \bibinfo {author} {\bibfnamefont {A.~D.}\
  \bibnamefont {Corso}}, \bibinfo {author} {\bibfnamefont {S.}~\bibnamefont
  {de~Gironcoli}}, \bibinfo {author} {\bibfnamefont {P.}~\bibnamefont
  {Delugas}}, \bibinfo {author} {\bibfnamefont {R.~A.~D.}\ \bibnamefont {Jr}},
  \bibinfo {author} {\bibfnamefont {A.}~\bibnamefont {Ferretti}}, \bibinfo
  {author} {\bibfnamefont {A.}~\bibnamefont {Floris}}, \bibinfo {author}
  {\bibfnamefont {G.}~\bibnamefont {Fratesi}}, \bibinfo {author} {\bibfnamefont
  {G.}~\bibnamefont {Fugallo}}, \bibinfo {author} {\bibfnamefont
  {R.}~\bibnamefont {Gebauer}}, \bibinfo {author} {\bibfnamefont
  {U.}~\bibnamefont {Gerstmann}}, \bibinfo {author} {\bibfnamefont
  {F.}~\bibnamefont {Giustino}}, \bibinfo {author} {\bibfnamefont
  {T.}~\bibnamefont {Gorni}}, \bibinfo {author} {\bibfnamefont
  {J.}~\bibnamefont {Jia}}, \bibinfo {author} {\bibfnamefont {M.}~\bibnamefont
  {Kawamura}}, \bibinfo {author} {\bibfnamefont {H.-Y.}\ \bibnamefont {Ko}},
  \bibinfo {author} {\bibfnamefont {A.}~\bibnamefont {Kokalj}}, \bibinfo
  {author} {\bibfnamefont {E.}~\bibnamefont {Küçükbenli}}, \bibinfo {author}
  {\bibfnamefont {M.}~\bibnamefont {Lazzeri}}, \bibinfo {author} {\bibfnamefont
  {M.}~\bibnamefont {Marsili}}, \bibinfo {author} {\bibfnamefont
  {N.}~\bibnamefont {Marzari}}, \bibinfo {author} {\bibfnamefont
  {F.}~\bibnamefont {Mauri}}, \bibinfo {author} {\bibfnamefont {N.~L.}\
  \bibnamefont {Nguyen}}, \bibinfo {author} {\bibfnamefont {H.-V.}\
  \bibnamefont {Nguyen}}, \bibinfo {author} {\bibfnamefont {A.~O.}\
  \bibnamefont {de-la Roza}}, \bibinfo {author} {\bibfnamefont
  {L.}~\bibnamefont {Paulatto}}, \bibinfo {author} {\bibfnamefont
  {S.}~\bibnamefont {Poncé}}, \bibinfo {author} {\bibfnamefont
  {D.}~\bibnamefont {Rocca}}, \bibinfo {author} {\bibfnamefont
  {R.}~\bibnamefont {Sabatini}}, \bibinfo {author} {\bibfnamefont
  {B.}~\bibnamefont {Santra}}, \bibinfo {author} {\bibfnamefont
  {M.}~\bibnamefont {Schlipf}}, \bibinfo {author} {\bibfnamefont {A.~P.}\
  \bibnamefont {Seitsonen}}, \bibinfo {author} {\bibfnamefont {A.}~\bibnamefont
  {Smogunov}}, \bibinfo {author} {\bibfnamefont {I.}~\bibnamefont {Timrov}},
  \bibinfo {author} {\bibfnamefont {T.}~\bibnamefont {Thonhauser}}, \bibinfo
  {author} {\bibfnamefont {P.}~\bibnamefont {Umari}}, \bibinfo {author}
  {\bibfnamefont {N.}~\bibnamefont {Vast}}, \bibinfo {author} {\bibfnamefont
  {X.}~\bibnamefont {Wu}},\ and\ \bibinfo {author} {\bibfnamefont
  {S.}~\bibnamefont {Baroni}},\ }\bibfield  {title} {\bibinfo {title} {Advanced
  capabilities for materials modelling with quantum espresso},\ }\href
  {http://stacks.iop.org/0953-8984/29/i=46/a=465901} {\bibfield  {journal}
  {\bibinfo  {journal} {Journal of Physics: Condensed Matter}\ }\textbf
  {\bibinfo {volume} {29}},\ \bibinfo {pages} {465901} (\bibinfo {year}
  {2017})}\BibitemShut {NoStop}%
\bibitem [{\citenamefont {Perdew}\ and\ \citenamefont
  {Zunger}(1981)}]{PerdewPRB1981}%
  \BibitemOpen
  \bibfield  {author} {\bibinfo {author} {\bibfnamefont {J.~P.}\ \bibnamefont
  {Perdew}}\ and\ \bibinfo {author} {\bibfnamefont {A.}~\bibnamefont
  {Zunger}},\ }\bibfield  {title} {\bibinfo {title} {Self-interaction
  correction to density-functional approximations for many-electron systems},\
  }\href {https://doi.org/10.1103/PhysRevB.23.5048} {\bibfield  {journal}
  {\bibinfo  {journal} {Phys. Rev. B}\ }\textbf {\bibinfo {volume} {23}},\
  \bibinfo {pages} {5048} (\bibinfo {year} {1981})}\BibitemShut {NoStop}%
\bibitem [{\citenamefont {Nagamatsu}\ \emph {et~al.}(2001)\citenamefont
  {Nagamatsu}, \citenamefont {Nakagawa}, \citenamefont {Muranaka},
  \citenamefont {Zenitani},\ and\ \citenamefont {Akimitsu}}]{NagamatsuNAT2001}%
  \BibitemOpen
  \bibfield  {author} {\bibinfo {author} {\bibfnamefont {J.}~\bibnamefont
  {Nagamatsu}}, \bibinfo {author} {\bibfnamefont {N.}~\bibnamefont {Nakagawa}},
  \bibinfo {author} {\bibfnamefont {T.}~\bibnamefont {Muranaka}}, \bibinfo
  {author} {\bibfnamefont {Y.}~\bibnamefont {Zenitani}},\ and\ \bibinfo
  {author} {\bibfnamefont {J.}~\bibnamefont {Akimitsu}},\ }\bibfield  {title}
  {\bibinfo {title} {Superconductivity at 39 k in magnesium diboride},\ }\href
  {https://doi.org/10.1038/35065039} {\bibfield  {journal} {\bibinfo  {journal}
  {Nature}\ }\textbf {\bibinfo {volume} {410}},\ \bibinfo {pages} {63}
  (\bibinfo {year} {2001})}\BibitemShut {NoStop}%
\bibitem [{\citenamefont {Eiguren}\ and\ \citenamefont
  {Ambrosch-Draxl}(2008{\natexlab{b}})}]{EigurenPRL2008}%
  \BibitemOpen
  \bibfield  {author} {\bibinfo {author} {\bibfnamefont {A.}~\bibnamefont
  {Eiguren}}\ and\ \bibinfo {author} {\bibfnamefont {C.}~\bibnamefont
  {Ambrosch-Draxl}},\ }\bibfield  {title} {\bibinfo {title} {Complex
  quasiparticle band structure induced by electron-phonon interaction: Band
  splitting in the
  $1\ifmmode\times\else\texttimes\fi{}1$$\mathrm{H}/\mathrm{W}(110)$ surface},\
  }\href {https://doi.org/10.1103/PhysRevLett.101.036402} {\bibfield  {journal}
  {\bibinfo  {journal} {Phys. Rev. Lett.}\ }\textbf {\bibinfo {volume} {101}},\
  \bibinfo {pages} {036402} (\bibinfo {year} {2008}{\natexlab{b}})}\BibitemShut
  {NoStop}%
\bibitem [{\citenamefont {Eiguren}\ \emph {et~al.}(2009)\citenamefont
  {Eiguren}, \citenamefont {Ambrosch-Draxl},\ and\ \citenamefont
  {Echenique}}]{EigurenPRB2009}%
  \BibitemOpen
  \bibfield  {author} {\bibinfo {author} {\bibfnamefont {A.}~\bibnamefont
  {Eiguren}}, \bibinfo {author} {\bibfnamefont {C.}~\bibnamefont
  {Ambrosch-Draxl}},\ and\ \bibinfo {author} {\bibfnamefont {P.~M.}\
  \bibnamefont {Echenique}},\ }\bibfield  {title} {\bibinfo {title}
  {Self-consistently renormalized quasiparticles under the electron-phonon
  interaction},\ }\href {https://doi.org/10.1103/PhysRevB.79.245103} {\bibfield
   {journal} {\bibinfo  {journal} {Phys. Rev. B}\ }\textbf {\bibinfo {volume}
  {79}},\ \bibinfo {pages} {245103} (\bibinfo {year} {2009})}\BibitemShut
  {NoStop}%
\bibitem [{\citenamefont {Garcia-Goiricelaya}\ \emph
  {et~al.}(2018)\citenamefont {Garcia-Goiricelaya}, \citenamefont {Gurtubay},\
  and\ \citenamefont {Eiguren}}]{GoiricelayaPRB2018}%
  \BibitemOpen
  \bibfield  {author} {\bibinfo {author} {\bibfnamefont {P.}~\bibnamefont
  {Garcia-Goiricelaya}}, \bibinfo {author} {\bibfnamefont {I.~G.}\ \bibnamefont
  {Gurtubay}},\ and\ \bibinfo {author} {\bibfnamefont {A.}~\bibnamefont
  {Eiguren}},\ }\bibfield  {title} {\bibinfo {title} {Coupled spin and
  electron-phonon interaction at the tl/si(111) surface from relativistic
  first-principles calculations},\ }\href
  {https://doi.org/10.1103/PhysRevB.97.201405} {\bibfield  {journal} {\bibinfo
  {journal} {Phys. Rev. B}\ }\textbf {\bibinfo {volume} {97}},\ \bibinfo
  {pages} {201405(R)} (\bibinfo {year} {2018})}\BibitemShut {NoStop}%
\bibitem [{\citenamefont {Choi}\ \emph {et~al.}(2002)\citenamefont {Choi},
  \citenamefont {Roundy}, \citenamefont {Sun}, \citenamefont {Cohen},\ and\
  \citenamefont {Louie}}]{ChoiPRB2002}%
  \BibitemOpen
  \bibfield  {author} {\bibinfo {author} {\bibfnamefont {H.~J.}\ \bibnamefont
  {Choi}}, \bibinfo {author} {\bibfnamefont {D.}~\bibnamefont {Roundy}},
  \bibinfo {author} {\bibfnamefont {H.}~\bibnamefont {Sun}}, \bibinfo {author}
  {\bibfnamefont {M.~L.}\ \bibnamefont {Cohen}},\ and\ \bibinfo {author}
  {\bibfnamefont {S.~G.}\ \bibnamefont {Louie}},\ }\bibfield  {title} {\bibinfo
  {title} {First-principles calculation of the superconducting transition in
  {M}g{B}$_{2}$ within the anisotropic {E}liashberg formalism},\ }\href
  {https://doi.org/10.1103/PhysRevB.66.020513} {\bibfield  {journal} {\bibinfo
  {journal} {Phys. Rev. B}\ }\textbf {\bibinfo {volume} {66}},\ \bibinfo
  {pages} {020513(R)} (\bibinfo {year} {2002})}\BibitemShut {NoStop}%
\bibitem [{\citenamefont {Li}\ \emph {et~al.}(2014)\citenamefont {Li},
  \citenamefont {Hao}, \citenamefont {Liu}, \citenamefont {Li},\ and\
  \citenamefont {Ma}}]{LiJCP2014}%
  \BibitemOpen
  \bibfield  {author} {\bibinfo {author} {\bibfnamefont {Y.}~\bibnamefont
  {Li}}, \bibinfo {author} {\bibfnamefont {J.}~\bibnamefont {Hao}}, \bibinfo
  {author} {\bibfnamefont {H.}~\bibnamefont {Liu}}, \bibinfo {author}
  {\bibfnamefont {Y.}~\bibnamefont {Li}},\ and\ \bibinfo {author}
  {\bibfnamefont {Y.}~\bibnamefont {Ma}},\ }\bibfield  {title} {\bibinfo
  {title} {The metallization and superconductivity of dense hydrogen sulfide},\
  }\href {https://doi.org/10.1063/1.4874158} {\bibfield  {journal} {\bibinfo
  {journal} {The Journal of Chemical Physics}\ }\textbf {\bibinfo {volume}
  {140}},\ \bibinfo {pages} {174712} (\bibinfo {year} {2014})},\ \Eprint
  {https://arxiv.org/abs/https://doi.org/10.1063/1.4874158}
  {https://doi.org/10.1063/1.4874158} \BibitemShut {NoStop}%
\bibitem [{\citenamefont {Duan}\ \emph {et~al.}(2014)\citenamefont {Duan},
  \citenamefont {Liu}, \citenamefont {Tian}, \citenamefont {Li}, \citenamefont
  {Huang}, \citenamefont {Zhao}, \citenamefont {Yu}, \citenamefont {Liu},
  \citenamefont {Tian},\ and\ \citenamefont {Cui}}]{DuanSCR2014}%
  \BibitemOpen
  \bibfield  {author} {\bibinfo {author} {\bibfnamefont {D.}~\bibnamefont
  {Duan}}, \bibinfo {author} {\bibfnamefont {Y.}~\bibnamefont {Liu}}, \bibinfo
  {author} {\bibfnamefont {F.}~\bibnamefont {Tian}}, \bibinfo {author}
  {\bibfnamefont {D.}~\bibnamefont {Li}}, \bibinfo {author} {\bibfnamefont
  {X.}~\bibnamefont {Huang}}, \bibinfo {author} {\bibfnamefont
  {Z.}~\bibnamefont {Zhao}}, \bibinfo {author} {\bibfnamefont {H.}~\bibnamefont
  {Yu}}, \bibinfo {author} {\bibfnamefont {B.}~\bibnamefont {Liu}}, \bibinfo
  {author} {\bibfnamefont {W.}~\bibnamefont {Tian}},\ and\ \bibinfo {author}
  {\bibfnamefont {T.}~\bibnamefont {Cui}},\ }\bibfield  {title} {\bibinfo
  {title} {Pressure-induced metallization of dense (h$_2$s)$_2$h$_2$ with
  high-tc superconductivity},\ }\href {https://doi.org/10.1038/srep06968}
  {\bibfield  {journal} {\bibinfo  {journal} {Scientific Reports}\ }\textbf
  {\bibinfo {volume} {4}},\ \bibinfo {pages} {6968} (\bibinfo {year}
  {2014})}\BibitemShut {NoStop}%
\bibitem [{\citenamefont {Drozdov}\ \emph {et~al.}(2015)\citenamefont
  {Drozdov}, \citenamefont {Eremets}, \citenamefont {Troyan}, \citenamefont
  {Ksenofontov},\ and\ \citenamefont {Shylin}}]{DrozdovNAT2015}%
  \BibitemOpen
  \bibfield  {author} {\bibinfo {author} {\bibfnamefont {A.~P.}\ \bibnamefont
  {Drozdov}}, \bibinfo {author} {\bibfnamefont {M.~I.}\ \bibnamefont
  {Eremets}}, \bibinfo {author} {\bibfnamefont {I.~A.}\ \bibnamefont {Troyan}},
  \bibinfo {author} {\bibfnamefont {V.}~\bibnamefont {Ksenofontov}},\ and\
  \bibinfo {author} {\bibfnamefont {S.~I.}\ \bibnamefont {Shylin}},\ }\bibfield
   {title} {\bibinfo {title} {Conventional superconductivity at 203 kelvin at
  high pressures in the sulfur hydride system},\ }\href
  {https://doi.org/10.1038/nature14964} {\bibfield  {journal} {\bibinfo
  {journal} {Nature}\ }\textbf {\bibinfo {volume} {525}},\ \bibinfo {pages}
  {73} (\bibinfo {year} {2015})}\BibitemShut {NoStop}%
\bibitem [{\citenamefont {Flores-Livas}\ \emph {et~al.}(2020)\citenamefont
  {Flores-Livas}, \citenamefont {Boeri}, \citenamefont {Sanna}, \citenamefont
  {Profeta}, \citenamefont {Arita},\ and\ \citenamefont
  {Eremets}}]{FloresLivasPHR2020}%
  \BibitemOpen
  \bibfield  {author} {\bibinfo {author} {\bibfnamefont {J.~A.}\ \bibnamefont
  {Flores-Livas}}, \bibinfo {author} {\bibfnamefont {L.}~\bibnamefont {Boeri}},
  \bibinfo {author} {\bibfnamefont {A.}~\bibnamefont {Sanna}}, \bibinfo
  {author} {\bibfnamefont {G.}~\bibnamefont {Profeta}}, \bibinfo {author}
  {\bibfnamefont {R.}~\bibnamefont {Arita}},\ and\ \bibinfo {author}
  {\bibfnamefont {M.}~\bibnamefont {Eremets}},\ }\bibfield  {title} {\bibinfo
  {title} {A perspective on conventional high-temperature superconductors at
  high pressure: Methods and materials},\ }\bibfield  {journal} {\bibinfo
  {journal} {Physics Reports}\ }\href
  {https://doi.org/https://doi.org/10.1016/j.physrep.2020.02.003}
  {https://doi.org/10.1016/j.physrep.2020.02.003} (\bibinfo {year}
  {2020})\BibitemShut {NoStop}%
\bibitem [{\citenamefont {Troyan}\ \emph {et~al.}(2020)\citenamefont {Troyan},
  \citenamefont {Semenok}, \citenamefont {Kvashnin}, \citenamefont {Sadakov},
  \citenamefont {Sobolevskiy}, \citenamefont {Pudalov}, \citenamefont
  {Ivanova}, \citenamefont {Prakapenka}, \citenamefont {Greenberg},
  \citenamefont {Gavriliuk}, \citenamefont {Struzhkin}, \citenamefont
  {Bergara}, \citenamefont {Errea}, \citenamefont {Bianco}, \citenamefont
  {Calandra}, \citenamefont {Mauri}, \citenamefont {Monacelli}, \citenamefont
  {Akashi},\ and\ \citenamefont {Oganov}}]{TroyanARXIV2019}%
  \BibitemOpen
  \bibfield  {author} {\bibinfo {author} {\bibfnamefont {I.~A.}\ \bibnamefont
  {Troyan}}, \bibinfo {author} {\bibfnamefont {D.~V.}\ \bibnamefont {Semenok}},
  \bibinfo {author} {\bibfnamefont {A.~G.}\ \bibnamefont {Kvashnin}}, \bibinfo
  {author} {\bibfnamefont {A.~V.}\ \bibnamefont {Sadakov}}, \bibinfo {author}
  {\bibfnamefont {O.~A.}\ \bibnamefont {Sobolevskiy}}, \bibinfo {author}
  {\bibfnamefont {V.~M.}\ \bibnamefont {Pudalov}}, \bibinfo {author}
  {\bibfnamefont {A.~G.}\ \bibnamefont {Ivanova}}, \bibinfo {author}
  {\bibfnamefont {V.~B.}\ \bibnamefont {Prakapenka}}, \bibinfo {author}
  {\bibfnamefont {E.}~\bibnamefont {Greenberg}}, \bibinfo {author}
  {\bibfnamefont {A.~G.}\ \bibnamefont {Gavriliuk}}, \bibinfo {author}
  {\bibfnamefont {V.~V.}\ \bibnamefont {Struzhkin}}, \bibinfo {author}
  {\bibfnamefont {A.}~\bibnamefont {Bergara}}, \bibinfo {author} {\bibfnamefont
  {I.}~\bibnamefont {Errea}}, \bibinfo {author} {\bibfnamefont
  {R.}~\bibnamefont {Bianco}}, \bibinfo {author} {\bibfnamefont
  {M.}~\bibnamefont {Calandra}}, \bibinfo {author} {\bibfnamefont
  {F.}~\bibnamefont {Mauri}}, \bibinfo {author} {\bibfnamefont
  {L.}~\bibnamefont {Monacelli}}, \bibinfo {author} {\bibfnamefont
  {R.}~\bibnamefont {Akashi}},\ and\ \bibinfo {author} {\bibfnamefont {A.~R.}\
  \bibnamefont {Oganov}},\ }\href@noop {} {\bibinfo {title} {Anomalous
  high-temperature superconductivity in {YH}$_6$}} (\bibinfo {year} {2020}),\
  \Eprint {https://arxiv.org/abs/1908.01534} {arXiv:1908.01534
  [cond-mat.supr-con]} \BibitemShut {NoStop}%
\bibitem [{\citenamefont {Kong}\ \emph {et~al.}(2019)\citenamefont {Kong},
  \citenamefont {Minkov}, \citenamefont {Kuzovnikov}, \citenamefont {Besedin},
  \citenamefont {Drozdov}, \citenamefont {Mozaffari}, \citenamefont {Balicas},
  \citenamefont {Balakirev}, \citenamefont {Prakapenka}, \citenamefont
  {Greenberg}, \citenamefont {Knyazev},\ and\ \citenamefont
  {Eremets}}]{KongARXIV2019}%
  \BibitemOpen
  \bibfield  {author} {\bibinfo {author} {\bibfnamefont {P.~P.}\ \bibnamefont
  {Kong}}, \bibinfo {author} {\bibfnamefont {V.~S.}\ \bibnamefont {Minkov}},
  \bibinfo {author} {\bibfnamefont {M.~A.}\ \bibnamefont {Kuzovnikov}},
  \bibinfo {author} {\bibfnamefont {S.~P.}\ \bibnamefont {Besedin}}, \bibinfo
  {author} {\bibfnamefont {A.~P.}\ \bibnamefont {Drozdov}}, \bibinfo {author}
  {\bibfnamefont {S.}~\bibnamefont {Mozaffari}}, \bibinfo {author}
  {\bibfnamefont {L.}~\bibnamefont {Balicas}}, \bibinfo {author} {\bibfnamefont
  {F.~F.}\ \bibnamefont {Balakirev}}, \bibinfo {author} {\bibfnamefont {V.~B.}\
  \bibnamefont {Prakapenka}}, \bibinfo {author} {\bibfnamefont
  {E.}~\bibnamefont {Greenberg}}, \bibinfo {author} {\bibfnamefont {D.~A.}\
  \bibnamefont {Knyazev}},\ and\ \bibinfo {author} {\bibfnamefont {M.~I.}\
  \bibnamefont {Eremets}},\ }\href@noop {} {\bibinfo {title} {Superconductivity
  up to 243 {K} in yttrium hydrides under high pressure}} (\bibinfo {year}
  {2019}),\ \Eprint {https://arxiv.org/abs/1909.10482} {arXiv:1909.10482
  [cond-mat.supr-con]} \BibitemShut {NoStop}%
\bibitem [{\citenamefont {Li}\ \emph {et~al.}(2015)\citenamefont {Li},
  \citenamefont {Hao}, \citenamefont {Liu}, \citenamefont {Tse}, \citenamefont
  {Wang},\ and\ \citenamefont {Ma}}]{LiSCR2015}%
  \BibitemOpen
  \bibfield  {author} {\bibinfo {author} {\bibfnamefont {Y.}~\bibnamefont
  {Li}}, \bibinfo {author} {\bibfnamefont {J.}~\bibnamefont {Hao}}, \bibinfo
  {author} {\bibfnamefont {H.}~\bibnamefont {Liu}}, \bibinfo {author}
  {\bibfnamefont {J.~S.}\ \bibnamefont {Tse}}, \bibinfo {author} {\bibfnamefont
  {Y.}~\bibnamefont {Wang}},\ and\ \bibinfo {author} {\bibfnamefont
  {Y.}~\bibnamefont {Ma}},\ }\bibfield  {title} {\bibinfo {title}
  {Pressure-stabilized superconductive yttrium hydrides},\ }\href
  {https://doi.org/10.1038/srep09948} {\bibfield  {journal} {\bibinfo
  {journal} {Scientific Reports}\ }\textbf {\bibinfo {volume} {5}},\ \bibinfo
  {pages} {9948} (\bibinfo {year} {2015})}\BibitemShut {NoStop}%
\bibitem [{\citenamefont {Peng}\ \emph {et~al.}(2017)\citenamefont {Peng},
  \citenamefont {Sun}, \citenamefont {Pickard}, \citenamefont {Needs},
  \citenamefont {Wu},\ and\ \citenamefont {Ma}}]{PengPRL2017}%
  \BibitemOpen
  \bibfield  {author} {\bibinfo {author} {\bibfnamefont {F.}~\bibnamefont
  {Peng}}, \bibinfo {author} {\bibfnamefont {Y.}~\bibnamefont {Sun}}, \bibinfo
  {author} {\bibfnamefont {C.~J.}\ \bibnamefont {Pickard}}, \bibinfo {author}
  {\bibfnamefont {R.~J.}\ \bibnamefont {Needs}}, \bibinfo {author}
  {\bibfnamefont {Q.}~\bibnamefont {Wu}},\ and\ \bibinfo {author}
  {\bibfnamefont {Y.}~\bibnamefont {Ma}},\ }\bibfield  {title} {\bibinfo
  {title} {Hydrogen clathrate structures in rare earth hydrides at high
  pressures: Possible route to room-temperature superconductivity},\ }\href
  {https://doi.org/10.1103/PhysRevLett.119.107001} {\bibfield  {journal}
  {\bibinfo  {journal} {Phys. Rev. Lett.}\ }\textbf {\bibinfo {volume} {119}},\
  \bibinfo {pages} {107001} (\bibinfo {year} {2017})}\BibitemShut {NoStop}%
\bibitem [{\citenamefont {Heil}\ \emph {et~al.}(2019)\citenamefont {Heil},
  \citenamefont {di~Cataldo}, \citenamefont {Bachelet},\ and\ \citenamefont
  {Boeri}}]{HeilPRB2019}%
  \BibitemOpen
  \bibfield  {author} {\bibinfo {author} {\bibfnamefont {C.}~\bibnamefont
  {Heil}}, \bibinfo {author} {\bibfnamefont {S.}~\bibnamefont {di~Cataldo}},
  \bibinfo {author} {\bibfnamefont {G.~B.}\ \bibnamefont {Bachelet}},\ and\
  \bibinfo {author} {\bibfnamefont {L.}~\bibnamefont {Boeri}},\ }\bibfield
  {title} {\bibinfo {title} {Superconductivity in sodalite-like yttrium hydride
  clathrates},\ }\href {https://doi.org/10.1103/PhysRevB.99.220502} {\bibfield
  {journal} {\bibinfo  {journal} {Phys. Rev. B}\ }\textbf {\bibinfo {volume}
  {99}},\ \bibinfo {pages} {220502(R)} (\bibinfo {year} {2019})}\BibitemShut
  {NoStop}%
\bibitem [{\citenamefont {Perdew}\ \emph {et~al.}(1996)\citenamefont {Perdew},
  \citenamefont {Burke},\ and\ \citenamefont {Ernzerhof}}]{PerdewPRL1996}%
  \BibitemOpen
  \bibfield  {author} {\bibinfo {author} {\bibfnamefont {J.~P.}\ \bibnamefont
  {Perdew}}, \bibinfo {author} {\bibfnamefont {K.}~\bibnamefont {Burke}},\ and\
  \bibinfo {author} {\bibfnamefont {M.}~\bibnamefont {Ernzerhof}},\ }\bibfield
  {title} {\bibinfo {title} {Generalized gradient approximation made simple},\
  }\href {https://doi.org/10.1103/PhysRevLett.77.3865} {\bibfield  {journal}
  {\bibinfo  {journal} {Phys. Rev. Lett.}\ }\textbf {\bibinfo {volume} {77}},\
  \bibinfo {pages} {3865} (\bibinfo {year} {1996})}\BibitemShut {NoStop}%
\bibitem [{\citenamefont {Hartwigsen}\ \emph {et~al.}(1998)\citenamefont
  {Hartwigsen}, \citenamefont {Goedecker},\ and\ \citenamefont
  {Hutter}}]{HartwigsenPRB1998}%
  \BibitemOpen
  \bibfield  {author} {\bibinfo {author} {\bibfnamefont {C.}~\bibnamefont
  {Hartwigsen}}, \bibinfo {author} {\bibfnamefont {S.}~\bibnamefont
  {Goedecker}},\ and\ \bibinfo {author} {\bibfnamefont {J.}~\bibnamefont
  {Hutter}},\ }\bibfield  {title} {\bibinfo {title} {Relativistic separable
  dual-space gaussian pseudopotentials from {H} to {R}n},\ }\href
  {https://doi.org/10.1103/PhysRevB.58.3641} {\bibfield  {journal} {\bibinfo
  {journal} {Phys. Rev. B}\ }\textbf {\bibinfo {volume} {58}},\ \bibinfo
  {pages} {3641} (\bibinfo {year} {1998})}\BibitemShut {NoStop}%
\bibitem [{\citenamefont {Goedecker}\ \emph {et~al.}(1996)\citenamefont
  {Goedecker}, \citenamefont {Teter},\ and\ \citenamefont
  {Hutter}}]{GoedeckerPRB1996}%
  \BibitemOpen
  \bibfield  {author} {\bibinfo {author} {\bibfnamefont {S.}~\bibnamefont
  {Goedecker}}, \bibinfo {author} {\bibfnamefont {M.}~\bibnamefont {Teter}},\
  and\ \bibinfo {author} {\bibfnamefont {J.}~\bibnamefont {Hutter}},\
  }\bibfield  {title} {\bibinfo {title} {Separable dual-space gaussian
  pseudopotentials},\ }\href {https://doi.org/10.1103/PhysRevB.54.1703}
  {\bibfield  {journal} {\bibinfo  {journal} {Phys. Rev. B}\ }\textbf {\bibinfo
  {volume} {54}},\ \bibinfo {pages} {1703} (\bibinfo {year}
  {1996})}\BibitemShut {NoStop}%
\bibitem [{\citenamefont {Lafuente-Bartolome}\ \emph
  {et~al.}(2020)\citenamefont {Lafuente-Bartolome}, \citenamefont {Gurtubay},\
  and\ \citenamefont {Eiguren}}]{AccompanyingPaper}%
  \BibitemOpen
  \bibfield  {author} {\bibinfo {author} {\bibfnamefont {J.}~\bibnamefont
  {Lafuente-Bartolome}}, \bibinfo {author} {\bibfnamefont {I.~G.}\ \bibnamefont
  {Gurtubay}},\ and\ \bibinfo {author} {\bibfnamefont {A.}~\bibnamefont
  {Eiguren}},\ }\bibfield  {title} {\bibinfo {title} {Fully anisotropic
  superconductivity with few {H}elmholtz {F}ermi-surface harmonics},\ }\href
  {https://doi.org/10.1103/PhysRevB.102.161107} {\bibfield  {journal} {\bibinfo
   {journal} {Phys. Rev. B}\ }\textbf {\bibinfo {volume} {102}},\ \bibinfo
  {pages} {161107} (\bibinfo {year} {2020})}\BibitemShut {NoStop}%
\bibitem [{\citenamefont {Xie}\ \emph {et~al.}(2019)\citenamefont {Xie},
  \citenamefont {Stewart}, \citenamefont {Hamlin}, \citenamefont {Hirschfeld},\
  and\ \citenamefont {Hennig}}]{XiePRB2019}%
  \BibitemOpen
  \bibfield  {author} {\bibinfo {author} {\bibfnamefont {S.~R.}\ \bibnamefont
  {Xie}}, \bibinfo {author} {\bibfnamefont {G.~R.}\ \bibnamefont {Stewart}},
  \bibinfo {author} {\bibfnamefont {J.~J.}\ \bibnamefont {Hamlin}}, \bibinfo
  {author} {\bibfnamefont {P.~J.}\ \bibnamefont {Hirschfeld}},\ and\ \bibinfo
  {author} {\bibfnamefont {R.~G.}\ \bibnamefont {Hennig}},\ }\bibfield  {title}
  {\bibinfo {title} {Functional form of the superconducting critical
  temperature from machine learning},\ }\href
  {https://doi.org/10.1103/PhysRevB.100.174513} {\bibfield  {journal} {\bibinfo
   {journal} {Phys. Rev. B}\ }\textbf {\bibinfo {volume} {100}},\ \bibinfo
  {pages} {174513} (\bibinfo {year} {2019})}\BibitemShut {NoStop}%
\end{thebibliography}%

\end{document}